\documentclass[a4paper,usenatbib,useAMS]{mnras}

\usepackage{graphicx}
\usepackage{lscape}
\usepackage{perpage}
\usepackage{amsmath}
\usepackage{color}
\MakePerPage{footnote}

\newcommand{\beq}	{\begin{equation}}
\newcommand{\eeq}	{\end{equation}}
\newcommand{\beqa}	{\begin{eqnarray}}
\newcommand{\eeqa}	{\end{eqnarray}}
\newcommand{\e}	        {$^{-1}$}
\newcommand{\ee}	{$^{-2}$}
\newcommand{\eee}	{$^{-3}$}

\newcommand{\calm}	{{\cal M}}

\newcommand\fnm		{\footnotemark}
\newcommand\fnt		{\footnotetext}

\newcommand{\alfven}    {{Alfv$\acute{\rm e}$n }}

\newcommand{\tf}	{t_f}

\newcommand{\va}	{v_{\rm A}}

\newcommand{\vrms}	{v_{\rm rms}}

\newcommand{\avir}      {\alpha_{\rm vir}}

\newcommand{\ma}	{{\calm_{\rm A}}}
\newcommand{\mao}	{{\calm_{\rm A,0}}}

\newcommand{\mmug}	{\mu{\rm G}}
\newcommand{\nbh}	{\bar n_{\rm H}}

\newcommand{\nbht}	{\bar n_{\rm H,\, 3}}

\newcommand{\sid}	{\sigma_{\rm 1D}}
\newcommand{\snt}       {\sigma_{\rm nt}}
\newcommand{\spc}	{\sigma_{\rm pc}}
\newcommand{\spcs}      {{\sigma_{\rm pc}^*}}

\newcommand{\NH}	{N_{\rm H}}
\newcommand{\tff}	{t_{\rm ff}}

\newcommand\cs		{c_{\rm s}}

\newcommand\pc		{{\rm pc}}

\def\tff				{t_{\rm ff}}

\title{Magnetized interstellar molecular clouds: II. The Large-Scale Structure and Dynamics of Filamentary Molecular Clouds}
\author[Pak Shing Li and Richard I. Klein]{Pak Shing Li$^{1}$\thanks{E-mail:psli@berkeley.edu (PSL)}, and Richard I. Klein$^{1,2}$\thanks{E-mail:klein@astron.berkeley.edu (RIK)} \\
$^{1}$Astronomy Department, University of California, Berkeley, CA 94720\\
$^{2}$Lawrence Livermore National Laboratory,P.O.Box 808, L-23, Livermore, CA 94550}
\begin{document}

\date{}

\pubyear{}

\label{firstpage}
\pagerange{\pageref{firstpage}--\pageref{lastpage}}
\maketitle

\begin{abstract}
We perform ideal MHD high resolution AMR simulations with driven turbulence and self-gravity and find that long filamentary molecular clouds are formed at the converging locations of large-scale turbulence flows and the filaments are bounded by gravity. The magnetic field helps shape and reinforce the long filamentary structures. The main filamentary cloud has a length of $\sim4.4$ pc. Instead of a monolithic cylindrical structure, the main cloud is shown to be a collection of fiber/web-like sub-structures similar to filamentary clouds such as L1495. Unless the line-of-sight is close to the mean field direction, the large-scale magnetic field and striations in the simulation are found roughly perpendicular to the long axis of the main cloud, similar to L1495. This provides strong support for a large-scale moderately strong magnetic field surrounding L1495. We find that the projection effect from observations can lead to incorrect interpretations of the true three-dimensional physical shape, size, and velocity structure of the clouds. Helical magnetic field structures found around filamentary clouds that are interpreted from Zeeman observations can be explained by a simple bending of the magnetic field that pierces through the cloud. We demonstrate that two dark clouds form a T-shape configuration which are strikingly similar to the Infrared dark cloud SDC13 leading to the interpretation that SDC13 results from a collision of two long filamentary clouds. We show that a moderately strong magnetic field ($\ma \sim 1$) is crucial for maintaining a long and slender filamentary cloud for a long period of time $\sim 0.5$ million years.
\end{abstract}

\begin{keywords}
ISM:magnetic fields, ISM:clouds, ISM:kinematics and dynamics, ISM:evolution, turbulence, methods:numerical
\end{keywords}

\section{Introduction}

Filamentary molecular clouds have received a great deal of attention recently. Massive and long filamentary clouds are commonly found inside Giant Molecular Clouds (GMCs) \citep{ber07,and14}, such as L1495 in the Taurus cloud complex \citep[e.g.][]{cha11}, the Serpens South cloud in the Serpens region \citep[e.g.][]{dha18}, the Musca molecular cloud \citep{kai16}, and the Orion A filament \citep{stu16}. Some filamentary clouds are dark in the mid-infrared wavelengths and belong to the category of Infrared Dark Clouds (IRDCs) with high column density ($N({\rm H}_2) > 10^{22}$ cm$^{-2}$). They are important in understanding massive star formation. Examples are IRDC G28.34+0.06 \citep{zha15} and G9.62+0.19 \citep{liu17}. The lengths of filamentary clouds can range from a few to more than ten parsecs (pcs) and have very high aspect ratio of more than ten.
There are also filamentary cloud systems that are composed of several very large filamentary clouds and appear to be connected, such as IC5146 \citep{arz11}, SDC13 \citep{per14}, and NGC1333 \citep{hac17}. There are  molecular dark clouds that show a complex web-like substructure, such as the Aquila region. This cloud appears to be a collection of many thin overlapped web-like filaments of the order of 0.1 pc in width \citep[e.g.][]{arz11,arz19}. Dense cores are found within these thin filaments or at their intersections \citep{kon15}. Our knowledge of filamentary molecular clouds has taken a large step forward with the launch of the Herschel space telescope \citep[e.g.]{and10,arz13,and14,kon15}. Recently, there are numerous ground-based observations focused on understanding the physical properties and dynamical structures of filamentary clouds. Using position-position velocity (PPV) data, velocity coherent fiber-like substructures have been identified in filamentary clouds, including L1495 \citep{hac13} and filamentary clouds in the Central Molecular Zone of our Galaxy \citep{hen16}. Some of them are described as braided because of the intertwined appearance of the substructures in the PPV data. From these high resolution observations, long filamentary clouds are shown to be not simple monolithic structures, but rather have complex web/fiber-like substructure.  Given this complexity, one could still assume a simple cylindrical geometry for the ease of discussion and analysis. Thus it remains a challenge to explain how the substructures found in filamentary clouds are formed \citep[e.g.][]{smi14,taf15}.

Filamentary structures are also commonly observed in numerical simulations at different scales in the study of star formation or at larger scales in galaxy disk modeling, regardless of their inclusion of magnetic fields in the simulations. For example, the hydrodynamics decay turbulence simulations by \citet{smi14} have no magnetic field included. However, they identify filamentary cloud structures, although the clouds appear to exhibit more clumpiness than a long coherent filament.
Nevertheless, complex velocity features in filamentary clouds are observable in hydrodynamics turbulence simulations \citep{moe15}.
Filamentary structures are also observed in disk galaxy simulations using Smooth Particle Hydrodynamics without magnetic fields \citep{dua16}. Due to limited resolution in \citet{dua16}, the identified massive filamentary structures are at the size scale of GMCs ($\ga 20$ pc). Simulating filamentary clouds in a highly supersonic turbulence system with magnetic fields is computationally expensive because of high \alfven velocities demanding a small time step. As a result, there are few MHD driven turbulence simulations of filamentary cloud formation of a large enough region ($> 4$ pc) and at the same time having high enough resolution ($< 0.01$ pc) to resolve substructures inside filamentary clouds. \citet{chi18} simulate a very large region and extract a (40 pc)$^3$ region to study filamentary clouds. The resolution is 0.1 pc, yet not enough to reveal substructures. They are able to identify a number of long filamentary clouds and focus on the study of fragmentation, instead of the physical properties of the filamentary cloud structure themselves. Their study provides no information on the magnetic field surrounding the clouds. \citet{fed16} performed a series of simulations in a (2 pc)$^3$ region, with resolution up to $2\times10^{-3}$ pc to investigate the effects of magnetic fields, gravity, and turbulence on the formation of filament structure within molecular clouds. He finds that in the ideal MHD and turbulence-only models, the thin filamentary structures commonly have a width of $\sim 0.1$ pc, as seen by some observations. However, \citet{pan17} point out that substructures of $\sim 0.1$ pc size could be the result of smoothing effects. Some recent observations of dark cloud substructures find filament substructures less than 0.1 pc \citep[e.g.][]{dha18,hac18}. There are no massive long filamentary clouds in \citet{fed16} MHD models similar to L1495 or the Musca cloud. 
This may be a result of the relatively weak magnetic field strength in these models (plasma $\beta$ of 0.33). The paper provides no information on the magnetic field structures of the filaments.

In this paper, we present results of the formation of long massive filamentary clouds from an ideal MHD, driven-turbulence simulation. The term {\it filament} is widely used in describing very large aspect ratio structures and, therefore describes both long filamentary clouds and their filamentary substructures. We use the term {\it filamentary cloud} for long ($\ga 3$ pc) and slender (aspect ratio $\ga 10$) clouds that have projected width $\ga 0.25$ pc. The mass per length is at least a few times the critical mass $2c_s^2/G$ and may be composed of thinner web/fiber-like substructures ($\la 0.1$ pc in width). The long and slender filamentary clouds formed in our simulation are suitable for comparison with observed filamentary dark clouds, such as L1495, and filamentary IRDCs, such as SDC13.
The main purpose of this paper is to provide a dynamically evolving scenario of the formation of filamentary clouds from the simulations, study their properties, and compare the simulation results with observations. Because of the full three-dimensional (3D) data available from our  simulations, we can reveal more features of the physical properties of filamentary clouds and the large scale environment surrounding the clouds than what is possible from the observations. We shall focus on the discussion of the large scale properties of the filamentary clouds and the region surrounding the clouds in this paper. The results of our simulations on the formation of web/fiber-like substructures and their dynamics will be reported in a future paper, Li, McKee, \& Klein, 2019 (paper III, in preparation). In Section 2, we discuss the simulation parameters and describe the two models involved in this work. In Section 3, we discuss the time evolving turbulence properties of the entire region of the initially strong field model undergoing gravitational collapse. In Section 4, we describe the global picture of the formation of the filamentary clouds in the simulated region through a video attached with this paper as supplementary material. This volume rendering of density helps to understand the large scale motion of the gas inside the turbulence region. The video elucidates the detailed discussions in the subsequent sections. In Section 5, 6, and 7, we discuss the physical properties of the filamentary clouds and the environment surrounding the clouds: Section 5 on density structure, Section 6 on magnetic field structure, and Section 7 on velocity field structure. Section 8 summarizes the physical properties of the main filamentary cloud in the simulation. Section 9 compares our simulation results with the dark cloud SDC13, which also has a T-shape spatial configuration similar to the two massive filamentary clouds in our simulation. In Section 10, we discuss the role of a strong magnetic field in creating and maintaining a long slender filamentary cloud by comparison with our weak field simulation. Finally, in Section 11, we summarize our conclusions.

\section{Simulation parameters and methods}
\label{sec:sim}

We use our multi-physics, adaptive mesh refinement (AMR) code \textsc{Orion2} \citep{li12a} to perform two large-scale simulations to study the formation and structure of filamentary clouds. \textsc{Orion2} is capable of solving ideal MHD along with coupled self-gravity \citep{mar08}, radiation transport and feedback physics. For the simulations reported in this paper, we study the formation of filamentary structures prior to the onset of star formation. Thus radiation transport and feedback physics are ignored. Our two simulations are basically ideal MHD driven turbulence simulations and gravity is turned on at a later time.
In the first crossing time, defined as $\tf \equiv \ell/v_{\rm rms}$, we drive the system at thermal Mach number $\calm = 10$ at the largest scales with wave number $k = 1 - 2$ on a single base grid of $512^3$ using the procedure described in \citet{mac99}. In the second crossing time we allow two levels of AMR with a refinement ratio of 2 at each fine level. When the refinement is less than 12 percent or significantly larger than 12 percent, we adjust the refinement requirement of pressure jumps (thermal pressure + magnetic pressure), density jump, and shear flows at the same time by the same amount during the second crossing time. The thresholds vary between 1.125 to 1.25 in the simulations. This ensures the high-mass end of the probability distribution function of density is equivalent to what could be reached in a $2048^3$ single-grid simulation as shown in \citet{li12a}. At the end of the second crossing time, we turn on gravity and continuously drive the system with a constant energy injection rate that maintains the entire system at Mach 10. We set the time $t=0$ at the moment when we turn on gravity.
After gravity is turned on, we include one additional refinement
requirement on the Jeans condition \citep{tru97} and we adopt a Jeans number of 1/8, which means that the Jeans length is at least resolved by 8 cells.
The finest resolution remains unchanged. As shown in the following section on scaling, the size of the entire region is $\ell = 4.55$ pc in correspondence to Mach 10,
enough for a $> 3\sim 4$ pc long filamentary clouds to form.
The highest resolution from our AMR simulations is $2.22 \times 10^{-3}$ pc, sufficient for the study of filamentary substructures of width at the order of 0.1 pc. We adopt periodic boundary conditions and assume an isothermal equation of state for the entire simulations.

Simulations of isothermal MHD turbulence are scale-free and are defined by two dimensionless parameters. They are the 3D thermal Mach number, $\calm = \vrms/c_s$, where $\vrms$ is the mass-weighted rms velocity, sound speed $c_s = 0.188$ km s$^{-1}$ at 10 K, and the \alfven Mach number $\ma = \vrms/\va$, where $\va$ is the mass-weighted \alfven velocity. Using the turbulent line-width-size relation \citep{mck07},
\beq
\snt=0.72\spcs R_{\rm pc}^{1/2}=0.51\spcs\ell_{\,\pc}^{1/2}~~~\mbox{km s\e},
\label{eq:lws}
\eeq
where $\snt$ is the mass-weighted 1D non-thermal velocity dispersion in a sphere of radius $R$, taken to be the same as in a box of size $\ell=2R$, and where
$\spcs \sim 1$ allows for deviations from the typical linewidth-size relation.  
The third dimensional parameter is the gravitational constant, $G$, together with the dimensionless virial parameter that measures the effect of self-gravity,
\beq
\avir\equiv \frac{5\sid^2 \ell}{2GM}\simeq \frac{5\snt^2 \ell}{2GM},
\label{eq:avir}
\eeq
where $\sid$ is the 1D mass-weighted velocity dispersion including the thermal velocity and where we have assumed that the flow is highly supersonic.

For highly supersonic, fully molecular gas, the physical parameters of the turbulent system---the size of the turbulent box, $\ell$,  
the flow time (or crossing time), $t_f$, the mass of the box, $M$, and the column density,
$\NH=\nbh\ell$---can
then be expressed as \citep[see the Appendix in][]{mck10}:
\begin{eqnarray}
\ell&=& \frac 23\;\frac{\calm^2 \cs^2}{(\spc^2/\mbox{ 1pc})}=0.0455\left(\frac{\calm^2 T_1}{\spcs^2}\right)~~\mbox{pc},
\label{eq:lscale}\\
t_f&=&\frac{\ell}{\vrms}=2.36\times 10^5\left(\frac{\calm T_1^{1/2}}{\spcs^2}\right)~~~\mbox{yr},
\label{eq:tscale}\\
\nbh&=&9.6\times 10^4\left(\frac{\spcs^4}{\avir\calm^2T_1}\right)~~~\mbox{cm\eee},
\label{eq:nscale}\\
M&=& 0.311\left(\frac{\calm^4T_1^2}{\avir\spcs^2}\right)~~M_\odot,
\label{eq:mscale}\\
\NH&=&1.34\times 10^{22}\left(\frac{\spcs^2}{\avir}\right)~~~\mbox{cm\ee}.
\label{eq:Nscale}
\end{eqnarray} 
The initial value of the uniform magnetic field is given by
\beq
B_0=4.56\;\left(\frac{\nbht T_1}{\beta_0}\right)^{1/2}\mmug
=31.6\left(\frac{\spcs^2}{\avir^{1/2}\ma}\right)\mmug.
\label{eq:bscale}
\eeq
where $\nbh$ is the average density of hydrogen nuclei, $\nbht=\nbh/(1000$~cm\eee), and $\beta_0=8\pi\rho \cs^2/B_0^2$ is the initial plasma beta.

The large-scale filamentary cloud simulation has $\calm=10$, $T=10$~K, and $\avir = \spcs=1$, so that the size of the simulated turbulent region is $\ell = 4.55$ pc, the mean number density $\nbh = 960$ cm$^{-3}$, the total mass $M = 3110 \,M_{\sun}$, and the mean column density of the system $\NH = 1.34\times 10^{22}$ cm$^{-2}$. The free-fall time of the entire turbulent region based on the mean density,
\beq
\tff=\left(\frac{3\pi}{32G\bar\rho}\right)^{1/2}=1.37\times 10^6\nbht^{-1/2}~~~\mbox{yr},
\label{eq:tff}
\eeq
is $\sim 1.4\times10^6$ yr, which is $0.59 t_{\rm f}$. The magnetic field strength in the simulation is moderately strong, 
with $\mao=1$, corresponding to an initial plasma $\beta_0 = 0.02$. The initial parameters of the simulation are summarized in Table \ref{tab:region}.

The relative importance of gravity and magnetic fields can be measured by the normalized mass-to-flux ratio, $\mu_\Phi=M/M_\Phi$, where the magnetic critical mass is given in terms of the magnetic flux threading the cloud, $\Phi$, by $M_\Phi=\Phi/(2\pi G^{1/2})$.
Gravity can overcome the field for $\mu_\Phi>1$, whereas gravitational collapse is impossible for $\mu_\Phi<1$ in the entire simulated region with a fixed mass and without turbulence.
For a cubical box, $\mu_\Phi$ is related to the other dimensionless parameters by $\mu_\Phi=(5\pi/6\avir)^{1/2}\mao=1.62$, which is only slightly supercritical. We have previously studied the magnetic field properties of the clumps in this simulation and reported the results in \citet{li15} (paper I).  We have performed an additional initially weak mean magnetic field simulation with $\mu_\Phi=16.2$
in that study for comparison. By comparing the observations \citep{li09,li11} on the differences between the magnetic field orientation of molecular clouds and the global mean field with our two models with different initial mean fields, we found that a moderately strong field model is necessary to explain the observed magnetic field orientation distribution.  In this paper, we show that the strong magnetic field is also important in forming and maintaining a long and slender filament by comparing these two models.

\begin{table}
\caption{Initial Physical Properties of the Entire Region}
\label{tab:region}
\begin{tabular}{lc}
\vspace{-0.3cm}\\
\hline
\hline
Thermal Mach number ($\calm$)  & 10              \\
\alfven Mach number ($\mao$)   & 1               \\
Temperature ($T$)              & 10 K            \\
Total mass ($M$)               & 3110 $M_{\odot}$\\
Mean density ($\bar{\rho}$)    & 960 cm$^{-3}$   \\
Size ($\ell$)                  & 4.55 pc         \\
Virial parameter ($\alpha_{\rm vir}$) & 1        \\
Plasma beta ($\beta_0$)        & 0.02            \\
Mass-to-flux ratio ($\mu_{\Phi}$)     & 1.62     \\
Flow time ($t_f$)             & $2.36\times10^6$ yr \\
Free-fall time ($t_{\rm ff}$)  & $1.4\times10^6$ yr \\
\hline
\end{tabular}
\end{table}

The simulation is run with an initially strong magnetic field for about $0.64\tff$ ($\sim 0.9 \times 10^6$ yrs), which is more than four times the free-fall time of the main filamentary cloud, but long enough to see the formation of filamentary clouds \citep{li18},
after the gravity is turned on.
The total computational cost of this simulation on 4096 processors is 1.8 million CPU hours, including the initial two crossing times driving without gravity.

\section{Turbulence properties of the system}
\label{sec:turbprop}

\subsection{Evolution of power spectra of velocity and density during gravitational collapse}
\label{sec:ps}

\begin{figure}
\includegraphics[scale=0.34]{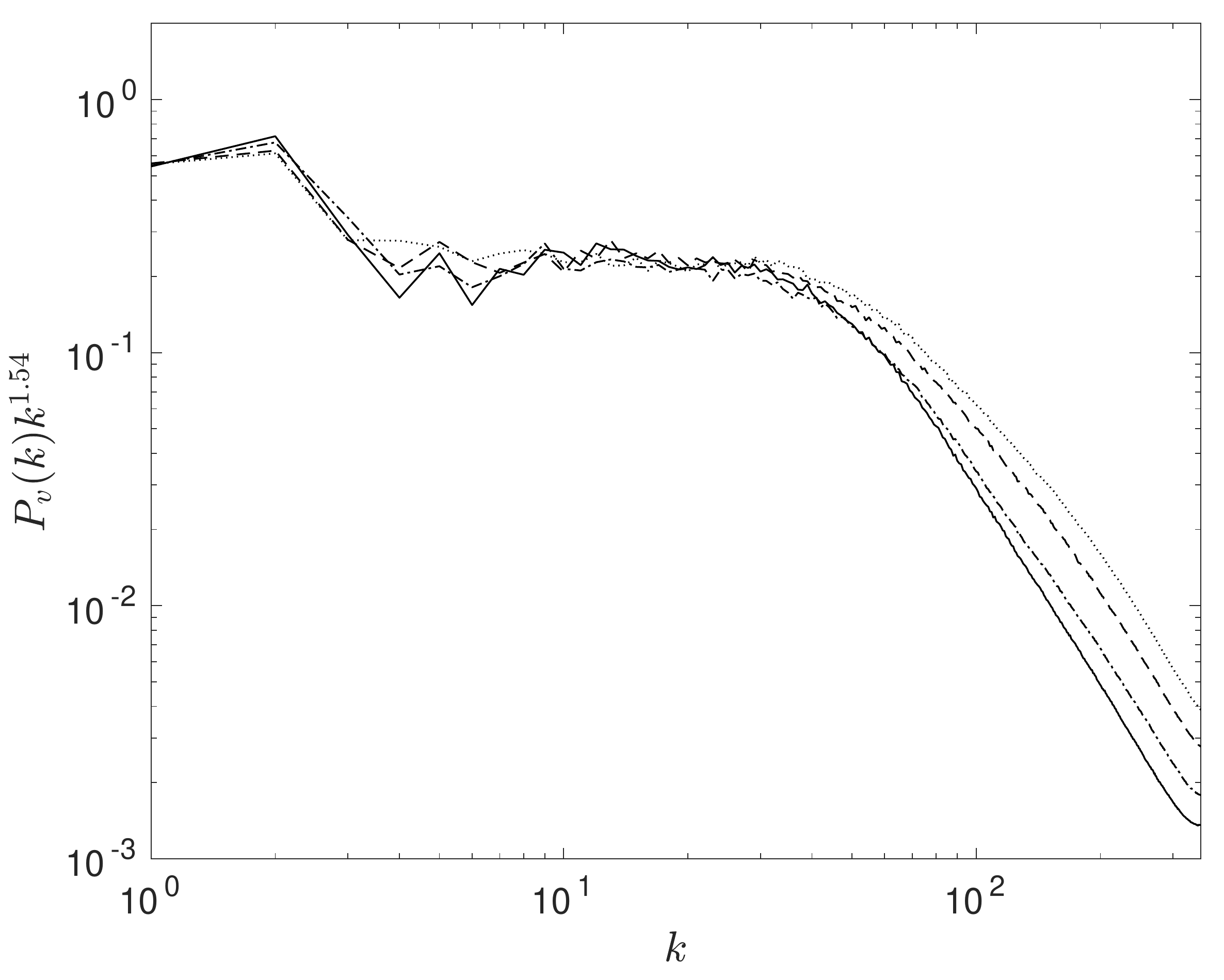}
\caption{Compensated velocity power spectra of the entire simulated region at time $t = 0, 0.3, 0.5,$ and $0.64 \tff$ (solid, dot-dashed, dashed, and dotted, respectively). The power spectra are normalized at $k = 0$ to be 1. The slopes of the inertial ranges remain about the same throughout the  simulation at a best fitted power slope of $\sim -1.54$. The power at high wave numbers slowly increases as the result of excited velocity dispersion due to gravitational collapse at smaller scales.
\label{fvps}}
\end{figure}

Fig. \ref{fvps} shows the compensated power spectra of velocity of the entire region at $t = 0, 0.3, 0.5,$ and $0.64 \tff$. During the gravitational collapse period, the system is continuously driven at constant injected kinetic energy that maintains a rms Mach 10 turbulent system. The inertial ranges extend to $k \sim 30$ and remain about the same throughout the entire simulation with a best fitted power slope of $\sim -1.54$. The length and apparent slope of the inertial range may be affected by the physically real bottleneck effect as the result of viscous suppression of small-scale modes of nonlinear interaction. The effect grows faster at smaller scales and leads to the pileup of energy in the inertial scales. Since the effect depends on how the dissipation scales with wavenumber \citep{fal94} and numerical viscosity depends on the numerical scheme being used, different codes will have a different degree of bottleneck effect \citep{kri11}.
The drop of power outside the inertial range is due to the limited resolution in the computational grid. The increase of power at smaller scales after gravity is turned on can still be seen indicating that turbulence at smaller scales will be enhanced in the process of gravitational collapse. 

\begin{figure}
\includegraphics[scale=0.33]{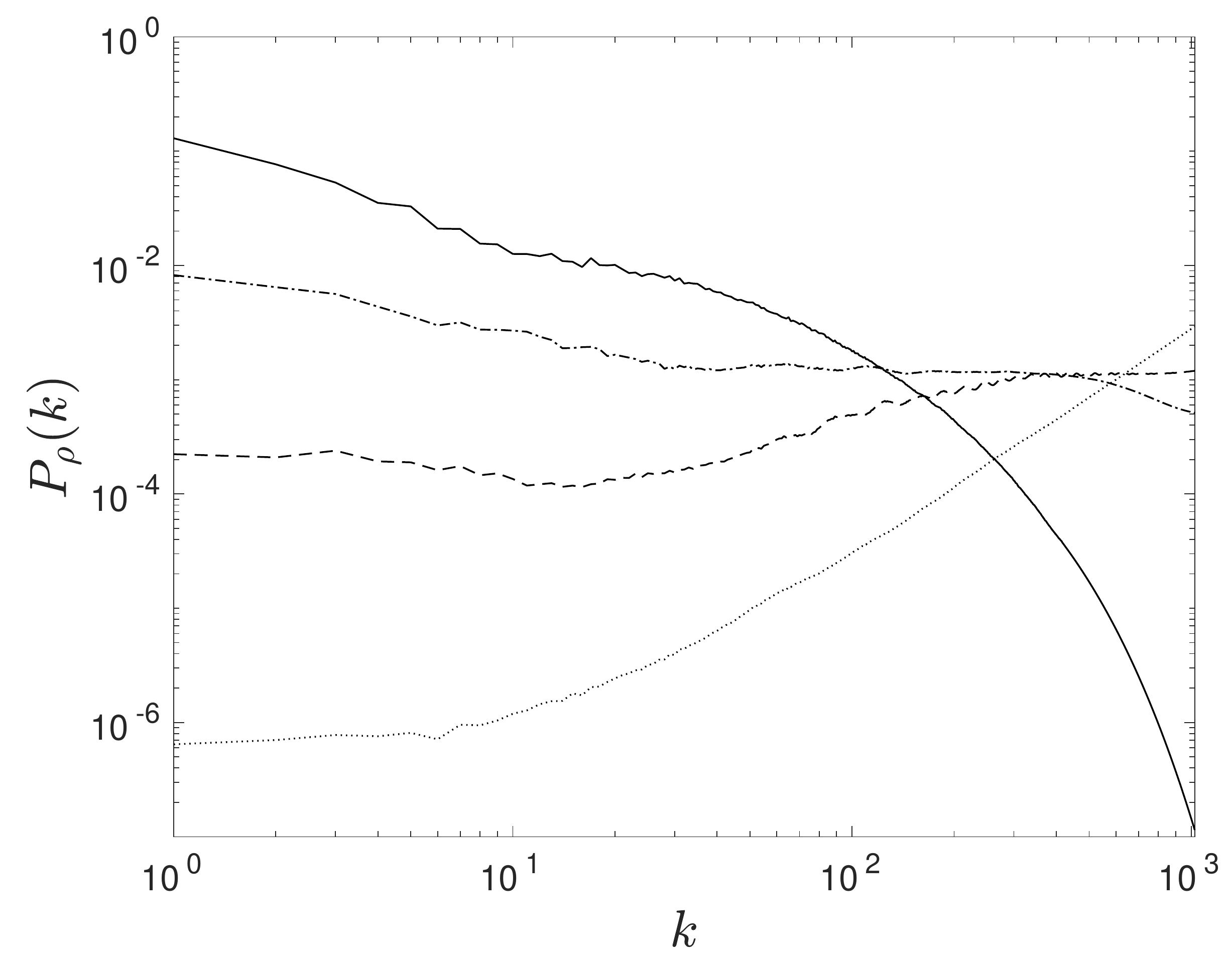}
\caption{Density power spectra of the entire simulated region at time $t = 0, 0.3, 0.5$, and $0.64 \tff$ (solid, dot-dashed, dashed, and dotted, respectively). The power spectra are normalized so that the total power of the spectra are the same.
\label{fdps}}
\end{figure}

Fig. \ref{fdps}, shows the density power spectra of the entire region at time $t = 0, 0.3, 0.5,$ and $0.64 \tff$. The power spectra are normalized so that the total power of the spectra are the same. The powers change drastically in time from negative to positive slopes. This is the result of the formation of dense structures at smaller scales during the gravitational collapse.

\subsection{Evolution of density PDF during gravitational collapse}
\label{sec:dpdf}

\begin{figure}
\includegraphics[scale=0.46]{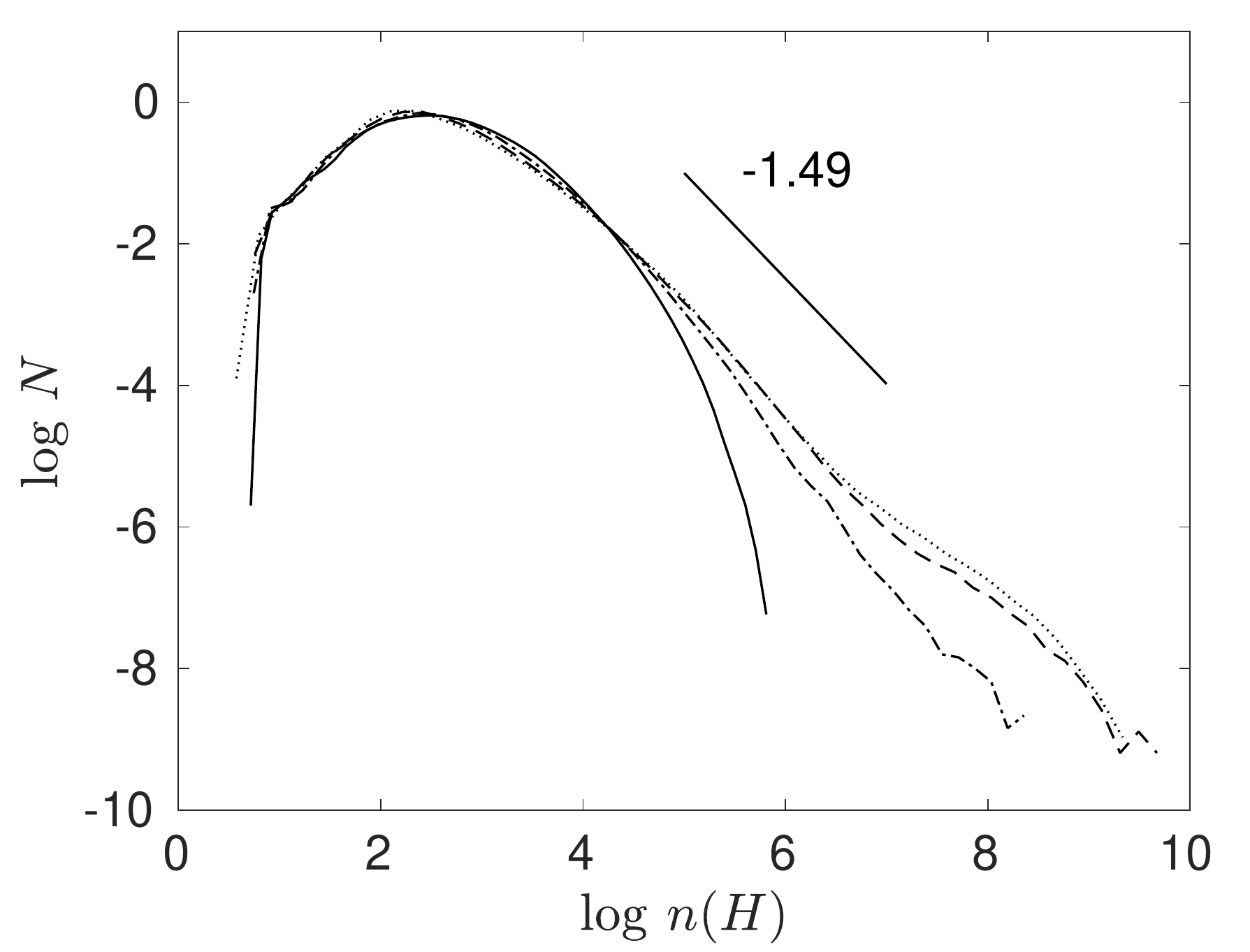}
\caption{Volume density PDFs of the entire simulated region at time $t = 0, 0.3, 0.5$, and $0.64 \tff$ (solid, dot-dashed, dashed, and dotted curves, respectively). The high density ends approach a power law distribution with index somewhat larger than -1.49 at the end of the simulation.
\label{dpdf}}
\end{figure}
In Fig. \ref{dpdf}, we show the volume density probability distribution functions (PDFs) of the entire volume at time $t = 0, 0.3, 0.5,$ and $0.64\tff$. The PDFs are normalized to have the same area under the curves. Just prior to switching on gravity, the density PDF is not symmetrical in the log scale, which is the result of using AMR in turbulence driving. Most of the low density regions are at the base level resolution. More dense cells are able to be refined at higher resolution. The non-monotonic low density end of the pdf is the result of a smaller amount of low density cells at the base level, where the sampling of the low density region is poor and creates the non-monotonic appearance of the PDF \citep[e.g.][]{kri06}.
The high density end of the distribution approaches a power law in time.
The power index is about -1.49 at the end of the simulation. This increase in power law index is again the result of increasingly dense structures formed in the clouds due to gravitational collapse.

\section{An overview of Filamentary Cloud formation in the simulation}
\label{sec:cloudform}
Before discussing the physical properties of the filamentary clouds in the simulation and comparison with observations, it is helpful to visualize the entire simulation. A video of volume rendering of density using visualization software VisIt \citep{chi12} is included for this purpose. The video lasts for about $0.9\times10^6$ years after gravity is turned on. The minimum density in the volume rendering is $2\times10^4$ g cm$^{-3}$. At the beginning of the video, which is after two crossing times of turbulence driving without gravity, as described in Section \ref{sec:sim}, a large amount of shock-compressed filamentary-like structures already exist and are distributed collectively in several large elongated clouds. 2D highly compressed thin slab structures are formed by plane-parallel shocks, mostly in rapid cooling environments such as molecular clouds near an isothermal state. With small perturbations, these 2D slab structures are subjected to the nonlinear thin-shell instability (NTSI) \citep[e.g.][]{vis94,kle98,mcl13} and do not remain stable for long. These structures corrugate and end up in filamentary-like structures. \citet{che17} provide a detail study of the NTSI in the presence of magnetic fields and they conclude that faint striations and more massive filament structures can be formed depending on the orientation of the velocity perturbation with respect to the magnetic field. In the video, we note that within two crossing times driving without gravity, these filamentary-like substructures are seen. The large scale magnetic field is aligned along the vertical direction. As time evolves, with the help of gravity, some existing filamentary-like structures are coalesced and form two prominent and denser filamentary clouds. The {\it main cloud} is roughly horizontal near the middle of the region and the {\it secondary cloud} is wider and less dense located behind the main cloud and extends to the top of the region. Shock-compressed filamentary-like structures are continuously forming throughout and are collected onto the main and secondary clouds. These two clouds become denser, thinner in size, and lengthen. At around 600,000 years, the secondary cloud collides with the main cloud at relative speed $\sim 1.5$ km s$^{-1}$ and fragments into 3 parts. The lowest portion merges with the main cloud. At this time, the main cloud is about 4.5 pc long.  An elongated and thin inverted-C shape gas stream moves down from the top onto the left side of the main cloud. At the end of the animation, the remaining secondary cloud evolves into clump-like gas clouds. The length and appearance of the main cloud remain about the same. Even though the main cloud is thin in size, complex filamentary substructures are clearly visible.

\begin{figure}
\includegraphics[scale=0.5]{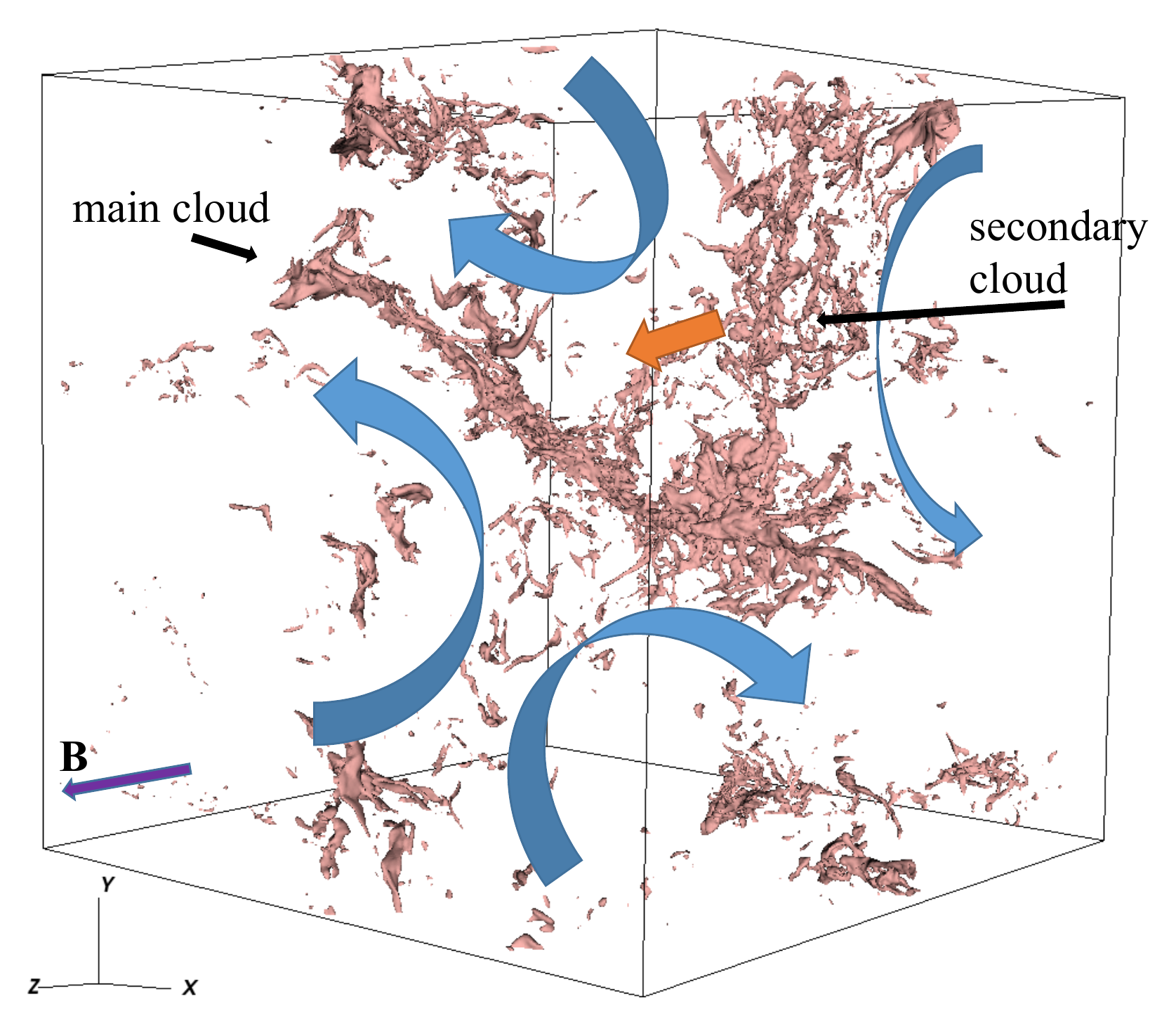}
\caption{Large scale velocity flow patterns around the main and secondary clouds in the simulation at time $t = 0.5 \tff$. The four blue color vectors indicate the large scale velocity flows converging at the location of the main cloud. The orange color vector shows the direction of the movement of the secondary cloud slowly approaching the main cloud and colliding with the main cloud from behind at around 600,000 years. The main magnetic field is along the z-axis.
\label{vfield3d}}
\end{figure}

The large scale movement of gas is mainly due to large scale turbulence driving. The large scale velocity flow directions are depicted in Figure \ref{vfield3d} at $0.5 \tff$.  Large scale turbulence eddies are created by the continuous turbulence driving throughout the simulation. It is clear that the main cloud is located at the convergence region of the four large eddies. In addition, the large circular flows not only deposit gas material onto the main cloud, but they also contribute to the stretching of the main cloud. In the video, the main cloud near the middle becomes very thin at about $0.46 \tff$. Due to the moderately strong magnetic field reinforcing the cloud and the continuous replenishment of gas from outside, the main cloud thickens again after this time.

\begin{figure}
\includegraphics[scale=0.73]{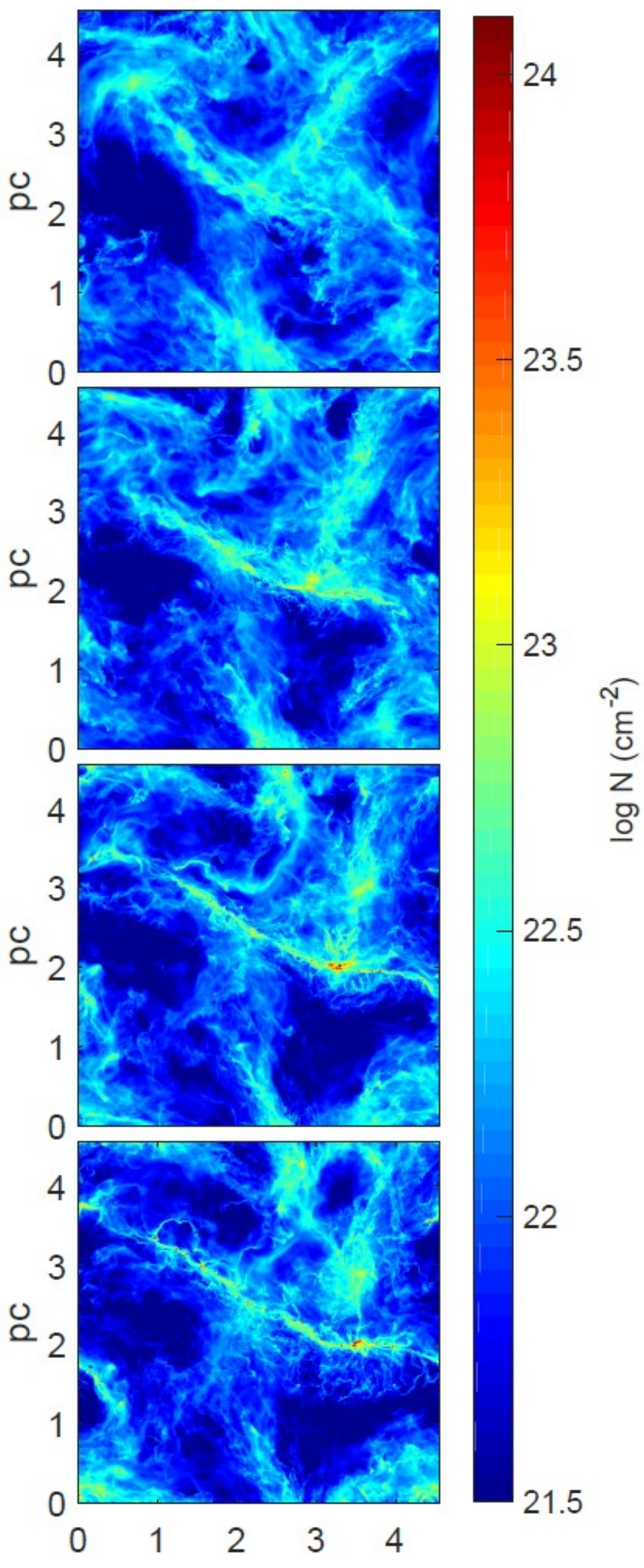}
\caption{Four snapshots of column density maps viewed along the mean magnetic field direction at time $t = 0, 0.3, 0.5,$ and $0.64 \tff$ from the top to bottom panels.
\label{Nevol}}
\end{figure}

In Fig. \ref{Nevol}, we show the column density maps of the entire region at time $t = 0, 0.3, 0.5,$ and $0.64 \tff$ for visual inspection. All of the maps are viewed along the mean field direction (z-axis). At $t = 0$, gas is distributed in several massive streams, but at much lower density. With gravity, shock-compressed material in the turbulence system is able to coalesce and help form the filamentary clouds. The existence of the main and secondary clouds are somewhat visible at t = 0 (top panel of Fig. \ref{Nevol}). The secondary cloud is more prominent at this time than the main cloud. Near the middle of the map the main cloud starts to form at about $0.3 \tff$. The secondary cloud is located near the right hand side of the main cloud and is oriented roughly perpendicular to the main cloud. These two clouds together form a T-shape configuration, similar to the structure in IRDC SDC13 \citep{per14}. More detailed comparison with SDC13 will be discussed in Section \ref{sec:sdc13}. The two filamentary clouds are in contact at around $t = 0.43 \tff$, instead of a false perception due to the viewing direction. As a result of the collision, additional protostars form in the $\sim 1$ pc long collision {\it junction}, later in time. We discuss this in detail in our study of protostellar cluster formation in \citet{li18}. The secondary cloud does not maintain its structure and disintegrates later. Part of it merges with the main cloud.  The main cloud condenses, lengthens, and becomes more prominent. In addition to the two large filamentary clouds, there are other gas clouds in the simulated region. The movement of gas in the entire simulated region is complex.

\begin{figure*}
\includegraphics[scale=0.7]{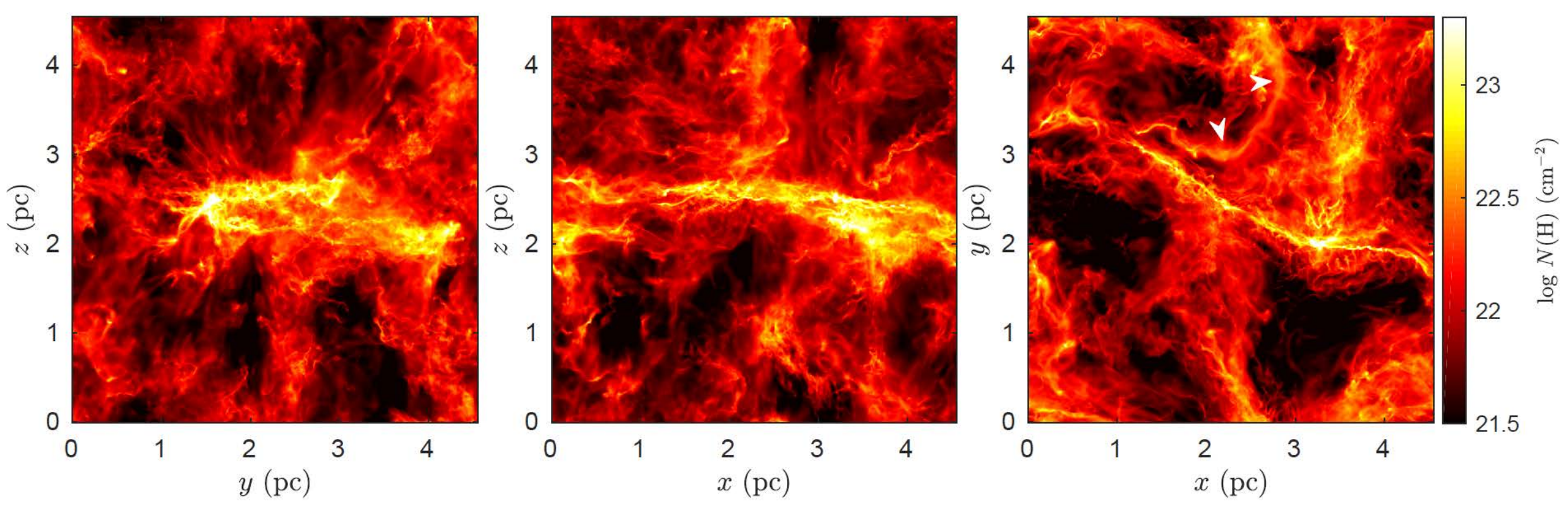}
\caption{Column density maps viewed along the three cardinal axes at $0.5 \tff$. The appearances of the clouds are so different that the two dense filamentary clouds in the T-shape configuration (right panel) could be treated as a single filamentary cloud (left and middle panels). The two white pointers in the right panel indicate a large low density reversed gas stream with the appearance of a reversed C-shape.
\label{Nmaps}}
\end{figure*}

\begin{figure*}
\includegraphics[scale=0.75]{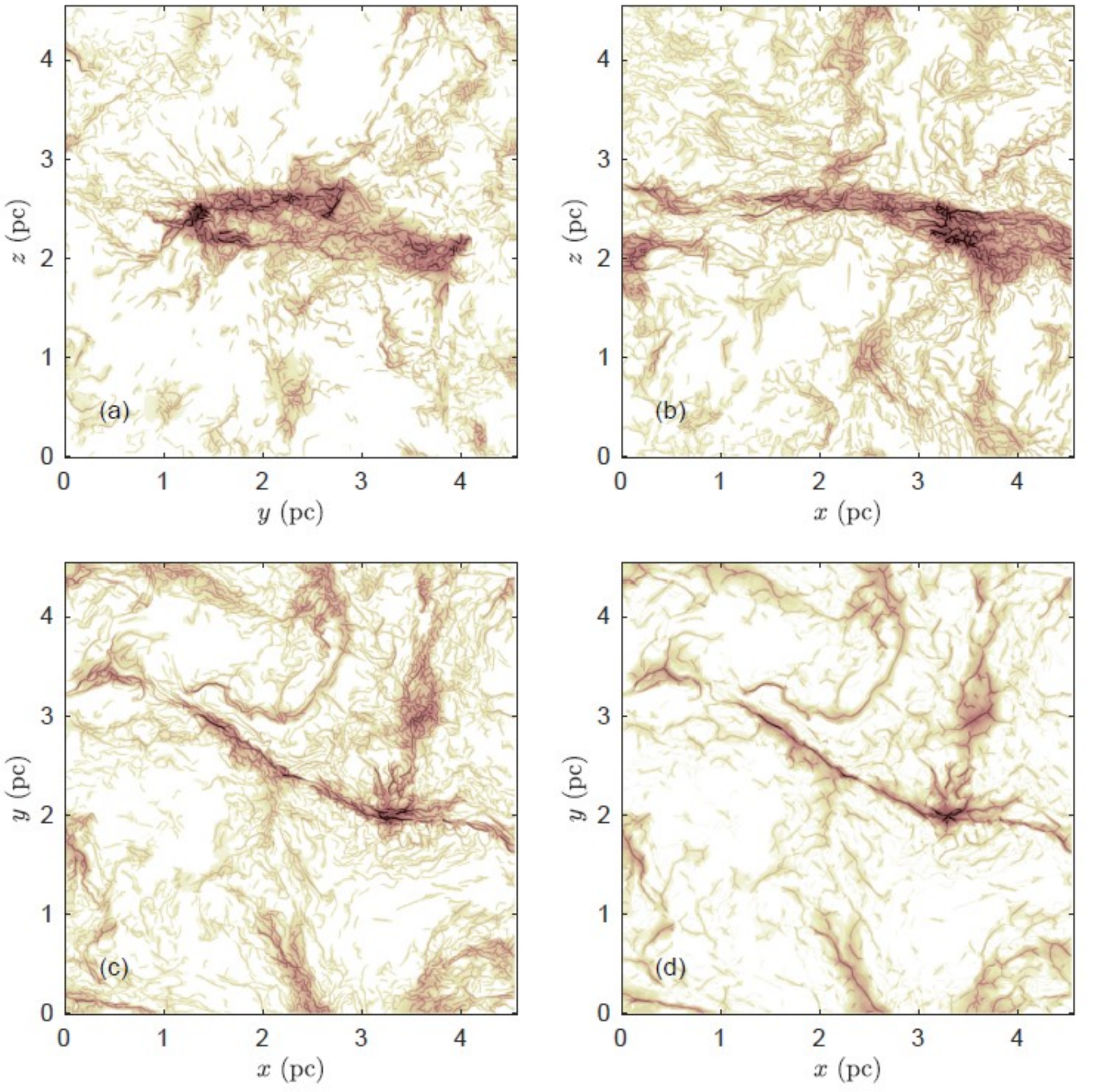}
\caption{Fiber/web like substructures of the dark clouds in the simulation region along the x-, y-, and z-axis are shown in panels (a) to (c). These substructures are identified using the \textsc{getfilament} algorithm. The panel (d) is the same as (c) but the resolution is ten times lower. At such lower resolution, the main cloud is identified as a single filamentary structure and most substructures are smoothed out.
\label{skeleton}}
\end{figure*}

As shown in Fig. \ref{Nevol}, the gas distribution at t = 0 is the result of large scale turbulence driving and the strength of the initial magnetic field. The effect from the magnetic field is discussed in Section \ref{sec:irdc_b3d}. Supersonic turbulence driving can both compress gas to very high density ($\propto \calm^2$) and destroy dense gas through shocks. The end result is a statistical equilibrium state. With gravity, the situation is different. Once the gas is brought together by converging flows and forms dense gas clouds, the clouds do not fragment easily because of gravitational binding unless the gas clouds encounter very strong shocks. With gravity, dense cloud clumps can continue to collapse and form dense cores. In our study of protostellar cluster formation in filamentary clouds based on this simulation \citep{li18}, there are 82 protostars formed at the end of the simulation at $0.5 \tff$. At later time, additional protostars will form in the cloud and some will become more massive. The protostellar feedback through radiation and powerful outflows eventually disrupt the structure of the main cloud. Therefore, in the following sections, we discuss the physical properties of the filamentary clouds using the simulation data up to $0.5 \tff$ ($\sim 700,000$ yrs). The comparison of the time evolving physical properties will be discussed at $t = 0, 0.3$, and $0.5 \tff$.

\section{Density structure}
\label{sec:irdc_d}

\subsection{Column density map}
\label{sec:irdc_c}
\begin{figure*}
\includegraphics[scale=0.7]{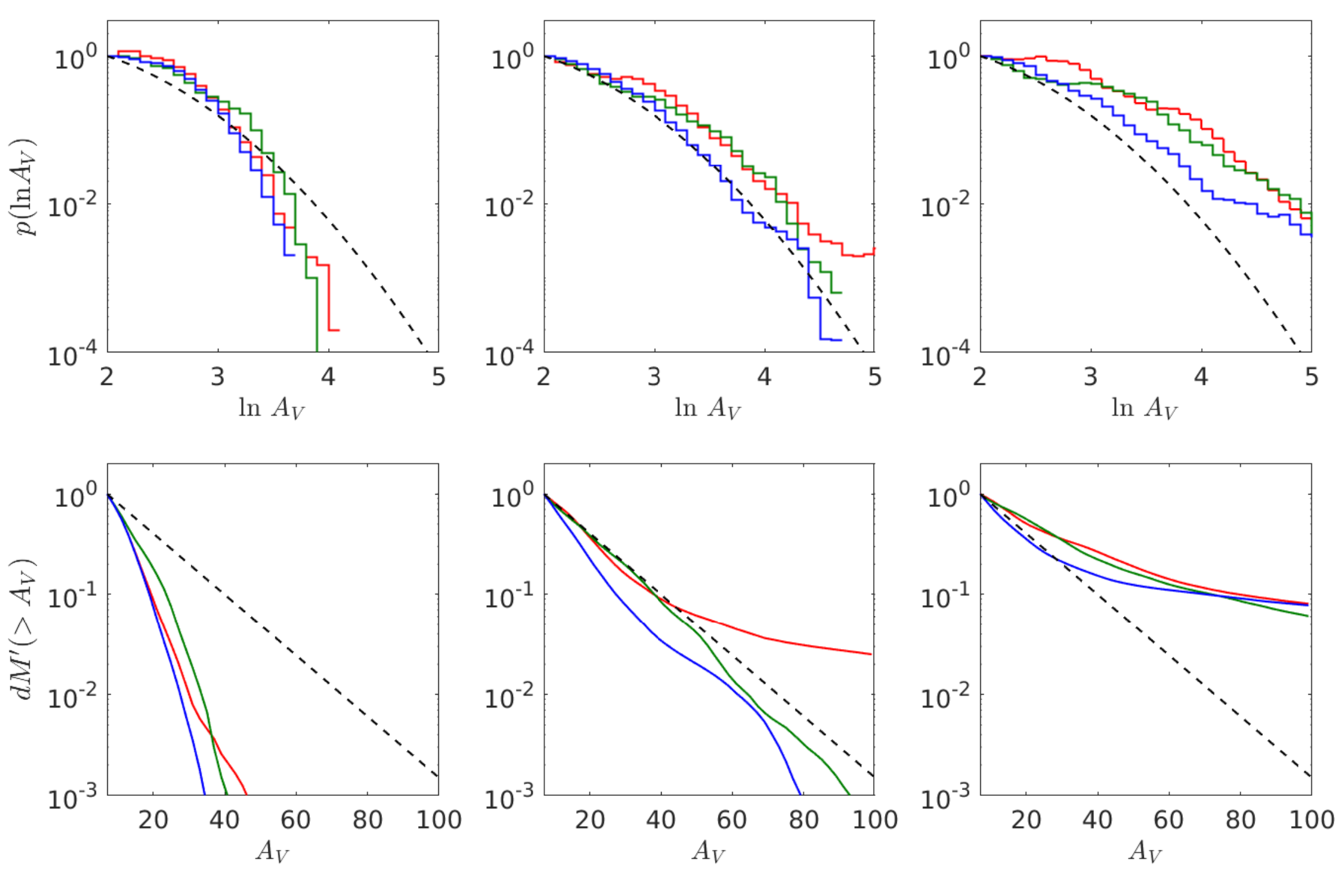}
\caption{Time evolution of column density PDFs (top rows) and CDFs (bottom rows) in term of extinction at different projections at time $t = 0, 0.3, 0.5\tff$ (left, middle, and right, respectively). The x-, y-, and z- projections are in red, deep green, and blue colors. The black dash curves in the top panels are a lognormal distribution with
$\mu=1.2$ and $\sigma_{\rm ln N}=0.9$, respectively. The black dash lines in bottom panels are an exponential function exp$(-0.07A_V)$. These dash curves and lines are the same as in \citet{kai13} figures 4 and 5 for visual comparison.
\label{Ncdf}}
\end{figure*}

In Fig. \ref{Nmaps}, the column density maps along the three cardinal axes at $0.5 \tff$ are plotted to illustrate how different the appearances of filamentary clouds are when looking from different directions. The true spatial configuration of the main and secondary clouds is better revealed in a T-shaped configuration in the projection along the z-axis (right panel). The appearance of the clouds is similar to the IRDC SDC 13 (see their figure 1 by rotating clockwise of $90^\circ$. However, if we have only the right map of Fig. \ref{Nmaps}, we do not know if the two filamentary clouds are actually spatially connected in this line-of-sight (los) direction. From the other two maps along the x- and y-axis (the left and middle panels respectively), the two clouds are in fact connected at this time. In the left and middle panel, the two clouds are so close on the x-y plane that the appearance is like a single filamentary cloud with a larger width. Therefore, the projection effect is important in the appearance of gas cloud. Line-of-sight velocity information may be helpful to distinguish the spatial structure of molecular clouds. Since the column densities of the filamentary clouds are larger than $2 \times 10^{22}$ cm$^{-2}$ no matter which direction we look at the system, the two filamentary clouds are also IRDCs.  In the right panel, the inverted-C shaped gas stream is above the main cloud and on the left of the secondary cloud. We note that because we are using periodic boundary conditions in the simulation, the left (or top) edge of the map is connected to the right (or bottom) edge of the map.

With careful visual examination, there are substructures inside the main and secondary clouds. We use a substructure identification algorithms based on morphology, such as \textsc{getfilament} \citep{men13}, to trace out the substructures. In Fig. \ref{skeleton}, we use \textsc{getfilament} to identify the thin fiber/web like substructures in the clouds along the three cardinal axes. The panel (d) is the same as the panel (c), but it is based on a ten times lower resolution column density map. With such low resolution, the substructures are largely smoothed out and the main cloud could be identified as a single filamentary structure. Thus high resolution observations can reveal rich substructures of dense molecular clouds. The fiber/web like substructures and their dynamics in the two clouds are discussed in paper III.

\subsection{Extinction PDFs and their time evolution}
\label{sec:irdc_e}

\begin{figure*}
\includegraphics[scale=0.55]{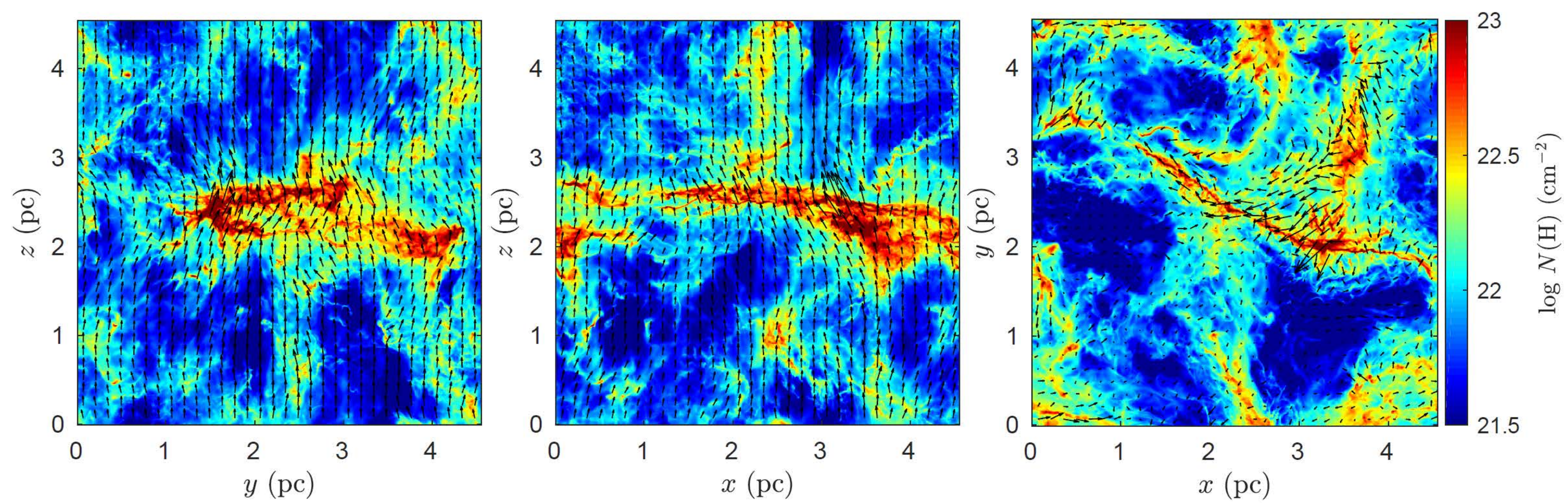}
\caption{The los density-weighted magnetic field, superimposed on top of the column density map, along the x-, y-, and z-axis (left to right, respectively) of the entire simulated region. The lengths of field vectors (at 0.14 pc pixel resolution) in the panels are scaled so that the longest vector in each panel corresponds to 105.7, 110.4, and 86.3 $\mu$G from the left to the right panel.
\label{bfield}}
\end{figure*}

Observations lack the spatial distribution information of gas along the los of a molecular cloud. Therefore, in order to compare simulations with observations, it is more appropriate to compare the column density distribution or extinction. For example, \citet{kai13} studied a number of dark clouds in Ophiuchus and found that some of the dark clouds show a log-normal distribution while others show a power law tail at the high density end. Different clouds also have different power law indexes. For the Ophiuchus clouds, the largest power law index is $\sim -1.15$ starting from extinction ln $A_V = 2$. Although a lognormal distribution is expected in a highly turbulent system, if a turbulent system is undergoing gravitational collapse, a power law tail at the high density end will develop as in Fig. \ref{dpdf} when high density structures are seen forming at smaller scales.
From the observations, the power law slope varies from about -4 to -1.5 \citep[e.g.][]{kai13,lin16,lin17}. For a direct comparison with the observational results in \citet{kai13}, we plot the column density PDFs in the unit of extinction magnitude, $A_V$, using the conversion between column density and extinction magnitude,
\beq
N_H = 1.9\times 10^{21} {\rm cm^2} \left( \frac{A_V}{mag} \right),
\eeq
as in \citet{kai13}. In Fig. \ref{Ncdf}, the top row panels show the PDFs of column densities along the three different cardinal projections at time $t=0, 0.3$, and $0.5\tff$ from left to right. The dashed curve in each panel is the same lognormal curve in the figure 4 of \citet{kai13} with the mean $\mu=1.2$ and the dispersion $\sigma_{\rm ln N}=0.9$. This curve serves as a reference to compare our figures with their figure 4. The bottom row of the panels show the column density cumulative distribution functions (CDFs) at the corresponding times. Note that the $A_V$ of the bottom row panels is plotted in linear scale, as in the figure 5 of \citet{kai13}. The dashed line is an exponential function, exp$(-0.07A_V)$, used in \citet{kai13} for visual comparison. From the observations, the column density PDFs and CDFs of molecular clouds have different shapes and slopes. Some are below and some are above the dashed reference curves or lines. The simulation shows that the high $A_V$ ends of the normalized density PDFs evolve from below the reference lognormal curves when gravity is absent at $t=0$, to a power law later in time. The CDFs in Fig. \ref{Ncdf} also show a clear transformation in appearance in time and occurs when denser structures are created in the dark clouds due to gravitational collapse. In addition, the CDF along the x-axis projection rises above the reference line earlier in time than the other two projections as a result of the projection effect. The change in both the power law slopes in PDFs and the shapes of CDFs can indicate different stages of gravitational collapse of molecular clouds.
\begin{figure*}
\includegraphics[scale=0.66]{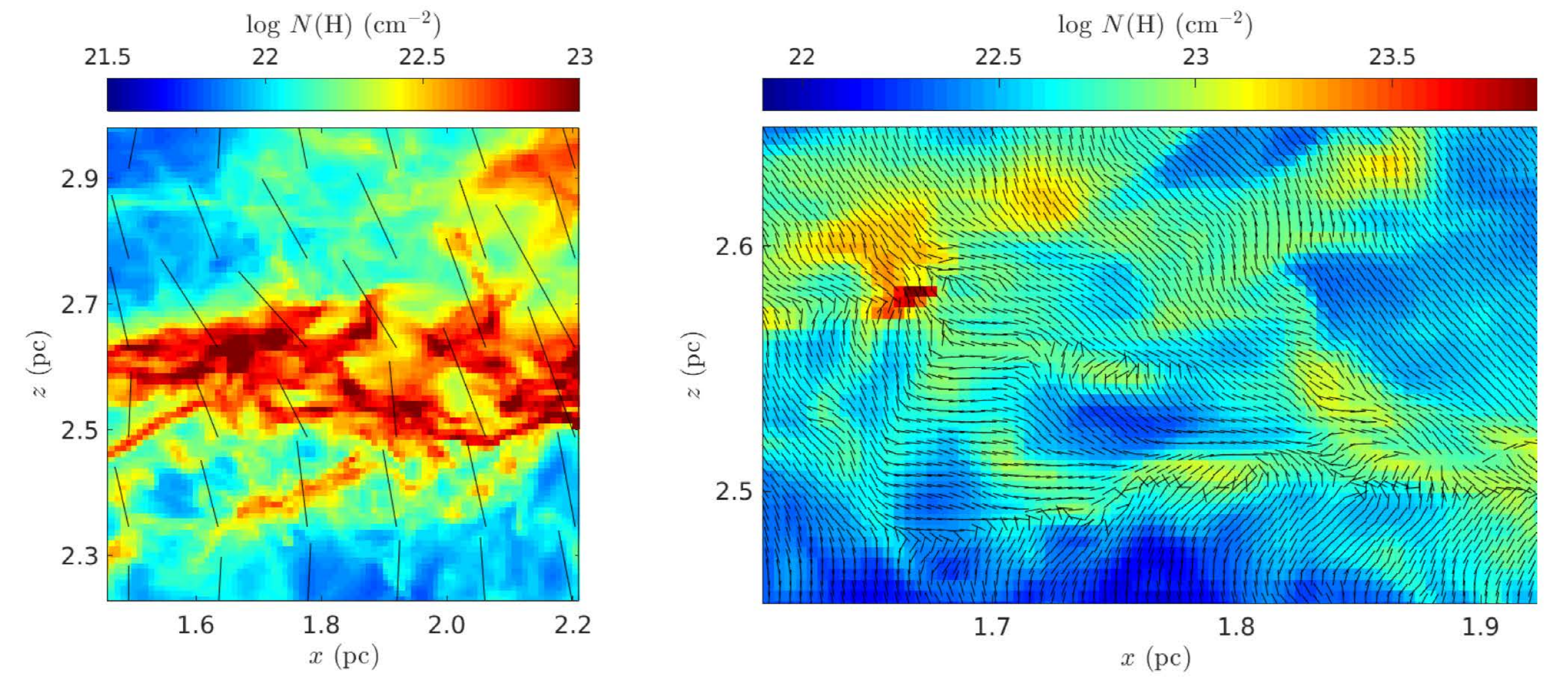}
\caption{(a) Zoom-in image of the main cloud in the middle panel in Fig. \ref{bfield}. The low resolution magnetic field at 0.14 pc resolution is roughly perpendicular to the main cloud's long axis. (b) Zoom-in image of panel (a) with high resolution magnetic field at $4.4\times10^{-3}$ pc resolution shows strong perturbations around dense substructures inside the main cloud.
\label{bfieldzoom}}
\end{figure*}
\begin{figure*}
\includegraphics[scale=0.8]{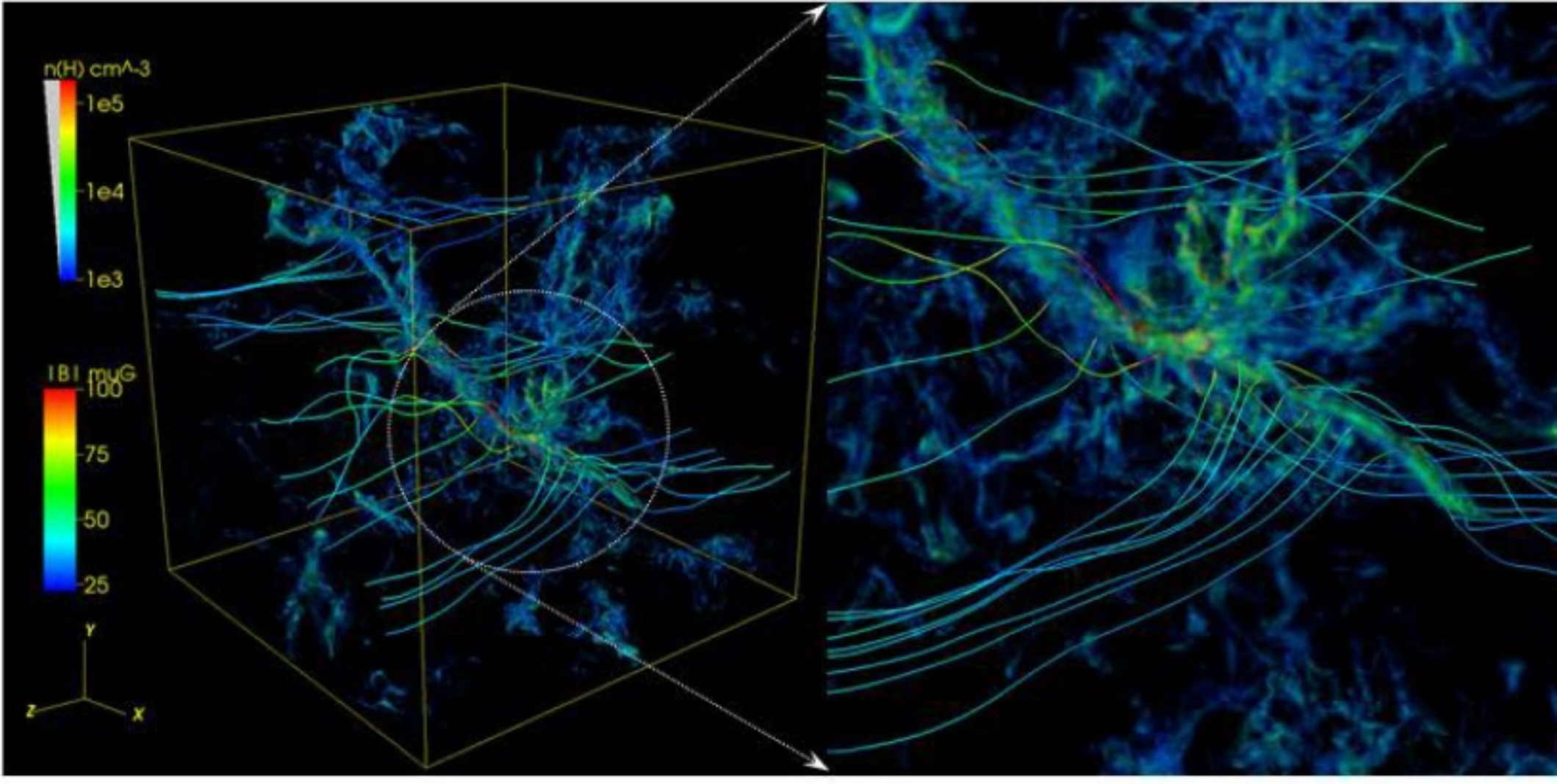}
\caption{Left panel: volume rendering of gas in the simulation region with magnetic field streamlines through the main cloud at a spacing of 0.25 pc. A larger deviation of the magnetic field from the mean field direction (z-axis) is seen near the main cloud. Right panel: a magnified view of a part of the main cloud (circled in on the left panel) with magnetic field streamlines spacing at 0.125 pc showing a systematic bending of magnetic field due to movement of the main cloud relative to the surrounding inter-cloud gas. The bending of the magnetic field can explain the Zeeman observations that are interpreted as a solenoidal field around filamentary clouds.
\label{bfield3d}}
\end{figure*}

\section{Magnetic field structure}
\label{sec:irdc_b}

\subsection{Projected and 3D magnetic field structures}
\label{sec:irdc_b3d}
The orientation of the magnetic field with respect to the long axis of a filamentary cloud is important. Modeling the magnetic field surrounding filamentary clouds is generally described by the combination of poloidal and toroidal fields along the cloud \citep[e.g.][]{fie00a,fie00b}, with a symmetrical axis along the cloud's long axis due to symmetry considerations. It is easier to set up magneto-hydrodynamic equilibrium initial conditions for the study of the stability and fragmentation of cylindrical filaments in this way \citep[e.g.][]{nag87,nak93,geh96,hos17}. Observations of polarization mapping usually give a more complex picture that results in different interpretations of the magnetic field structure surrounding filamentary clouds.  An example is a helical structure wrapping around the clouds \citep[e.g.][]{hei87,mat02,tah18} and a simple magnetic field penetrating the cloud roughly normal to the long axis \citep[e.g.][]{pal13}. The argument suggesting a helical field structure is taken from Zeeman observations showing the field is found to be pointing "into" the map along one side of the filament and pointing "out of" the map along the opposite side of the filament. There are also polarization observations indicating that the magnetic field appears to be roughly perpendicular to the cloud's long axis \citep[e.g.][]{cha11}. Strong support for this perpendicular orientation scenario is from the Herschel view of the large scale faint striations around the B211/B213/L1495 region in Taurus that matches well with the magnetic field orientation obtained from polarization mapping around L1495. Recent numerical and observational studies have found that for low density filamentary gas the magnetic field tends to be parallel to the filament's long axis. For massive filamentary gas, such as in filamentary dark clouds, the magnetic field tends to be perpendicular to the cloud's long axis \citep[e.g.][]{cha11,cox16,pla16,liu18,juv18}. \citet{rei18} examine the observables that are sensitive in determining the magnetic field morphology of molecular clouds and find that both dust polarization and Zeeman splitting are required in understanding the 3D field configuration. In this section, we discuss the magnetic field structure around the filamentary cloud in our simulation.

Fig. \ref{bfield} shows the density-weighted magnetic field projected onto the plane of the sky along the three cardinal axes. Because of the limited visible resolution in a figure, the magnetic field is smoothed to $0.14$ pc pixel size, close to the Planck observational resolution.
For the left and middle panels, projected along the x- and y-axis, the vertical axis is the direction of the mean magnetic field. Initially, with $\ma = 1$, the z-component of the magnetic field dominates the other two orthogonal components on the maps. However, there are still some large deviations of field vectors away from the mean field direction at some locations. From these two viewing directions, we conclude that the projected large scale magnetic field is roughly perpendicular to the long axis of the main cloud. In the right panel, we are looking along the mean field direction. The appearance of the projected field on the map is essentially chaotic.
The projected field, mainly excited by large scale turbulence, results in different angles with the long axis of the main cloud.
The lengths of the vectors in the three panels are scaled differently (see figure caption) to enhance more visible vectors. If the Davis-Chandrasekhar-Fermi (DCF) method is used to estimate the magnetic field strength on the plane of the entire map, the mean field strength on the plane of the right panel map is found to be very small, which is correct because the mean field is along the los in this map.
The mean field strength on the plane of the sky can be quite different from the true 3D field strength. If the right panel of Fig. \ref{bfield} is used, a very different conclusion from the other two panels will be drawn such that the magnetic field in this region is weak and the magnetic field is not as perpendicular to the long axis of the main cloud.

\begin{figure}
\includegraphics[scale=0.65]{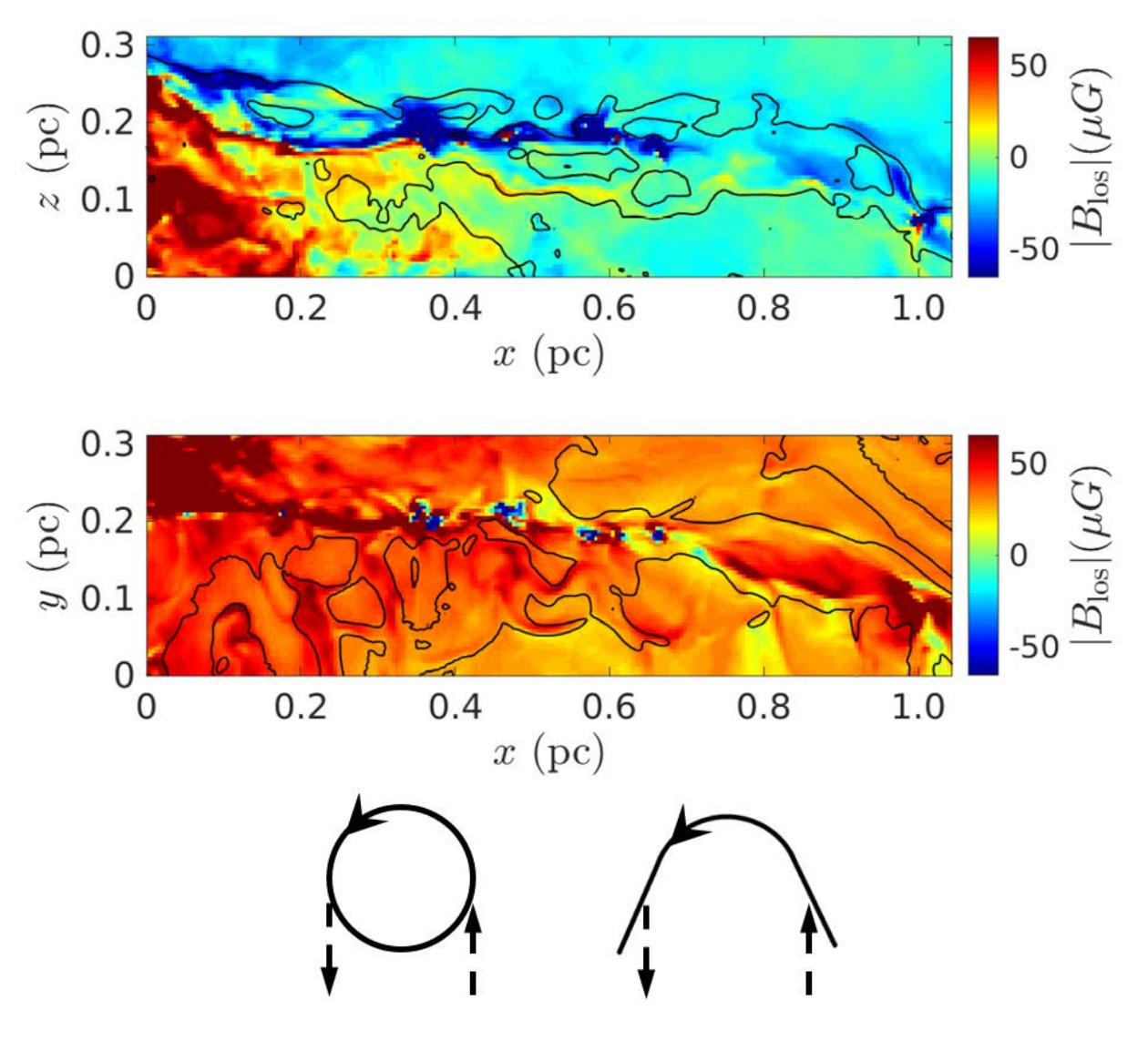}
\caption{Top: Density-weighted magnetic field map viewing along the y-axis. The contour is at a column density of $N(H) = 4\times10^{22}$ cm$^{-2}$, outlining the portion of the main cloud with a large systematic bending shown in Fig. \ref{bfield3d}. The magnetic field is pointing toward the observer (positive values) along the bottom half of the filament and away from the observer (negative values) along the top half of the filament, as seen in \citet{hei87} Zeeman observations. Middle: Same as top panel but viewing along the z-axis. The contour is at column density of $2\times10^{22}$ cm$^{-2}$. In this direction, observers will see the magnetic field pointing predominantly in one direction (positive value) throughout the filamentary cloud except for some small scale reversal of field directions near the dense cores. Bottom: the Sketches show that the opposite los magnetic field components at the two edges of the main cloud measured by the observers appear the same from a helical field and from a bending field.
\label{zeeman}}
\end{figure}

Fig. \ref{bfieldzoom} shows the magnetic field morphology inside the main cloud. We enlarge a small portion ($\sim0.75$ pc) of the main cloud in the middle panel of Fig. \ref{bfield} in the left panel of Fig. \ref{bfieldzoom}. In this map, there are many dense and thin filamentary substructures. The large scale low resolution ($\sim 0.071$ pc) field orientation is mostly perpendicular to the cloud's long axis. On the right panel, a smaller portion of the left panel ($\sim 0.3$ pc long) is enlarged and the magnetic field is at the resolution of $4.4\times10^{-3}$ pc. In this high resolution map, the magnetic field morphology is more complex. Magnetic fields are strongly distorted around dense structures including the fiber-like substructures and cores. The strong deviation of the magnetic field orientation from the mean field, as large as $90^{\circ}$, is the result of gravitational collapse, enhanced turbulence, and gas flow along the cloud's long axis (see Section \ref{sec:irdc_lv}). A change of field orientation inside filamentary clouds is also predicted in the study by \citet{gom18}
and from observational hints in L1495 \citep{and17}. The detailed analysis of the interplay between gravity, turbulence, and the magnetic field inside the main cloud will be reported in paper III.

\begin{figure}
\includegraphics[scale=0.4]{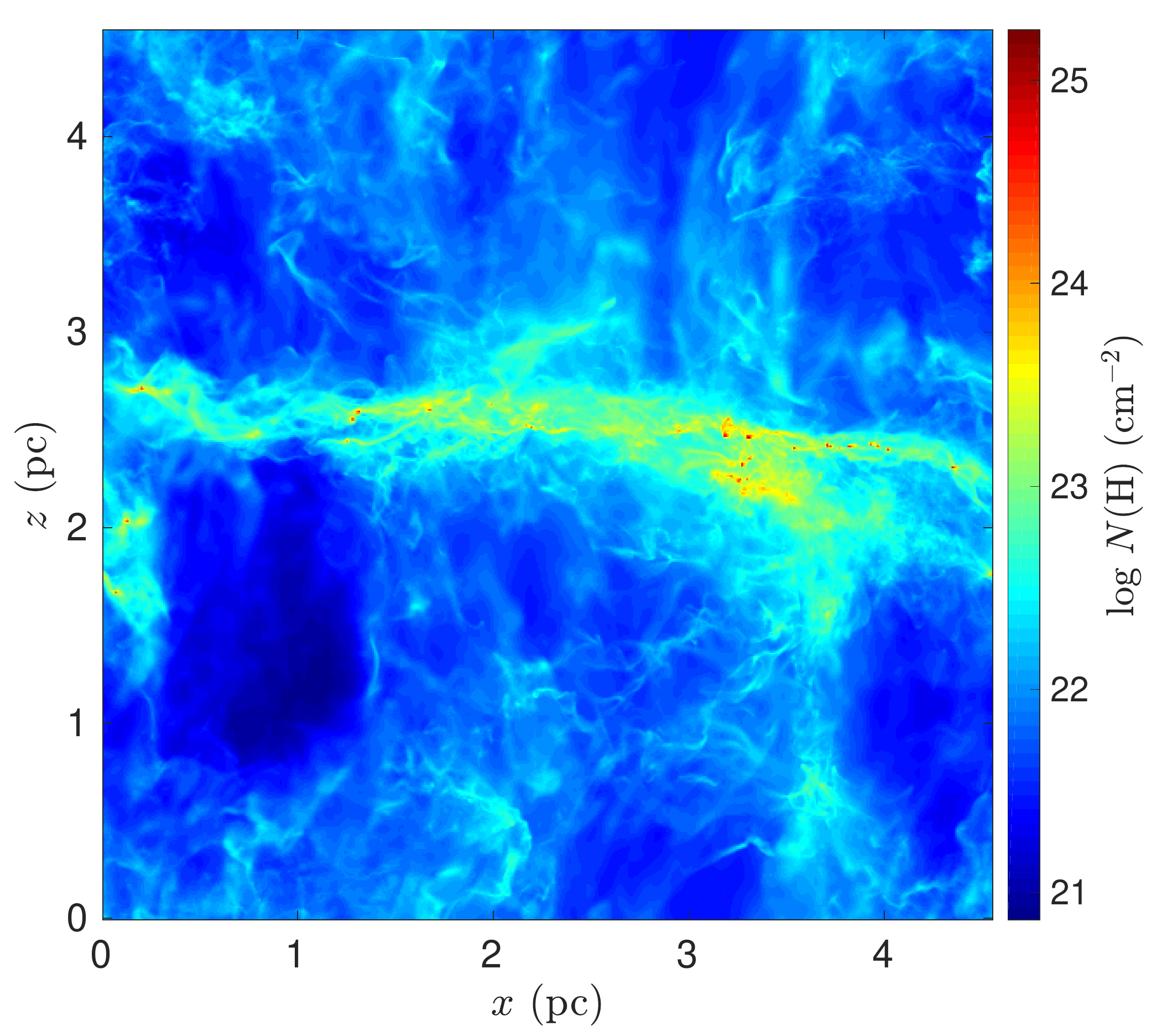}
\caption{Top panel: Column density map integrated over a depth of 2.28 pc, just covering the entire main cloud along the y-axis. The faint striations are clear above and below  the main cloud and are roughly along the vertical axis.
\label{striations}}
\end{figure}

\begin{figure*}
\includegraphics[scale=0.75]{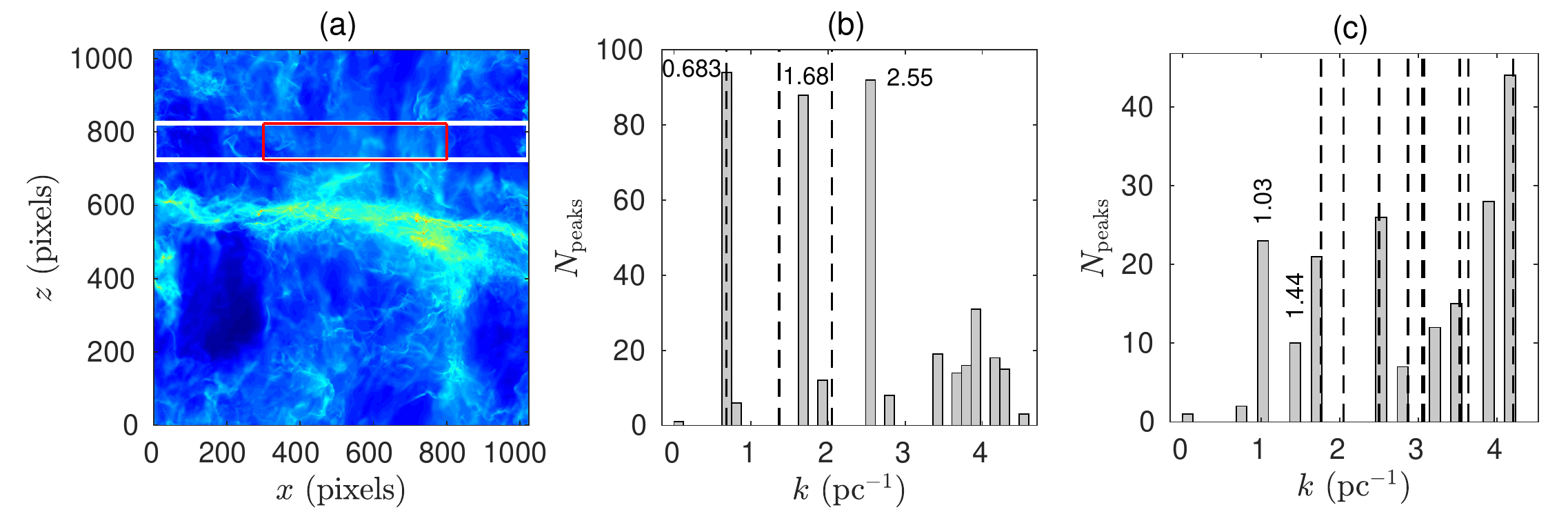}
\caption{(a) A $100 \times 1024$ pixel stripe (white) that covers the entire width of the simulated region and a shorter $100 \times 500$ pixel stripe (red) that covers only the visible striations are used for analysis of normal modes of magnetosonic waves. (b) The peak counts of normal modes inside the white-colored long stripe. The main peaks are roughly pointing to the existence of a true filamentary cloud. The three dashed lines corresponds to modes (1,0), (2,0), and (3,0) using the spatial frequency of the first peak. (c) The peak counts of normal modes inside the shorter red colored stripe. Most of the peaks fit well with the low frequency normal modes and incorrectly suggest a rectangular shaped cloud of $\sim 3.05 \times 2.18$ pc$^2$. See Section \ref{sec:irdc_striation} for discussion.
\label{striations_comp}}
\end{figure*}

To get a better visualization of the 3D magnetic field structure, we create a volume rendering of the dark clouds in the simulation with magnetic field streamlines piercing through the main cloud at roughly equal spacing of 0.25 pc along the long axis as shown in the left panel of Fig. \ref{bfield3d}. In this figure, the magnetic fields are mostly aligned close to the z-axis but with larger distortion near the cloud. The large scale magnetic field has not been perturbed significantly as is expected in a turbulent system with a moderately strong initial mean field of $\ma = 1$. Stronger distortion to the field occurs near the dense dark clouds. This is the result of a magnetic field frozen with the dense gas that has larger kinetic energy than the low density inter-cloud gas. At one end of the main cloud, we note a systematic bending of the magnetic field, which is the result of the movement of that part of the main cloud relative to the inter-cloud gas. That part of the cloud is magnified and is shown in the right panel of Fig. \ref{bfield3d}, with streamlines plotted at closer spacing of $\sim 0.125$ pc. Small scale, large distortion of the magnetic field around the dense cores is not possible to visualize here because of the choice of resolution in the figure. There is no clear evidence of well defined large scale helical-type magnetic fields surrounding the main cloud as suggested by \citet{hei87}. Zeeman observations of a dark cloud in the Orion region \citep{hei87} indicate that the magnetic field "goes in" along one edge of the filament cloud and "comes out" at the other edge. A helical structure is suggested to explain the observations. Using  rotational measurement data, \citet{tah18} also suggest similar "one side in and other side out" magnetic fields along the los at two sides of several clouds, although the measurement data is insufficient to make definite conclusions. If the cloud has a magnetic field structure similar to the picture in Fig. \ref{bfield3d} and we are looking close to the direction perpendicular to the large scale mean field direction (along y-axis in this case), we obtain a similar Zeeman observation result as in \citet{hei87} because the bending of the magnetic field creates a magnetic field component pointing toward the observer along one edge of the cloud and away from the observer along the other edge (cf. Fig. \ref{zeeman}). In our case, the mean magnetic field is strong. The cloud must have a very large angular momentum to overcome the magnetic tension in order to coil up the magnetic field to form a helical structure. Therefore, the scenario in Fig. \ref{bfield3d} of the bending of the magnetic field due to gas flow around filament clouds would be more likely to explain the Zeeman observation results in \citet{hei87} in the case of a strong mean magnetic field.

\subsection{Striations}
\label{sec:irdc_striation}

Striations are low density, faint sub-filamentary structures that appear perpendicular to the main filamentary clouds, as seen in Musca molecular cloud \citep{cox16} and in Taurus B211/B213/L1495 region \citep{gol08,pal13}. Striation structures are commonly seen in turbulence simulations with a moderately strong magnetic field \citep[e.g.][]{li12b}.
\citet{che17} suggest that striations are created as a result of corrugation of an oblique shock-compressed 2D thin slab of gas along the magnetic field direction due to the NTSI. As predicted in \citet{che17}, striations will be seen when $\calm \ma \la 0.2$. In our case $\calm \ma \sim 0.1$ and we expect to see striations in our simulation. It is seen that there are very faint vertical structures above and below the main cloud in the left and middle panels in Fig. \ref{Nmaps}. The mean field is along the vertical axis. If we integrate only the region enclosing the main cloud instead of the entire simulated region, the striations can be better revealed. In Fig. \ref{striations}, we show a column density map of the middle panel of Fig. \ref{Nmaps} integrated from 1.25 to 3.53 pc measuring from the bottom edge of the right panel along y-axis in Fig. \ref{Nmaps}. The faint striations are now clear and they are mostly along the mean field direction. This implies that there is a large scale moderately strong magnetic field roughly perpendicular to the long filament cloud L1495 in Taurus as proposed by \citet{pal13}. With more gas in the front and at the back of the main cloud, the striations are less clear in Fig. \ref{Nmaps}. This demonstrates that striations may be common, but sometimes unclear or invisible because of projection overlapping. Note that we do not see any striations in the right panel of Fig. \ref{Nmaps}.

\citet{tri18} attempt to use the normal modes of striations around the Musca molecular cloud to reveal the dimension of the cloud along the los, based on the assumption that striations are the result of magnetosonic waves excited from the \alfven waves originating from the molecular cloud during the gravitational collapse. By selecting a thin region away from the Musca molecular cloud covering the striations in the image, they perform a spectral analysis using the Lamb-Scargle periodogram technique \citep{sca82,tow10} on the data in the stripe. They find that the second smallest spatial frequency of the normal mode (n,m),
\beq
k_{nm} = \sqrt{\big( \frac{\pi n}{L_x} \big)^2+\big( \frac{\pi m}{L_y} \big)^2}
\label{mode}
\eeq
is not a multiple of the smallest spatial frequency, but is between $k(1,0)$ and $k(2,0)$. Note that the spatial size in units of pc is the inverse of the spatial frequency and multiplied by $\pi$. If the cloud is truly a filamentary cloud such that the orthogonal dimension along the los is much less than the length of the cloud, the second smallest spatial frequency is expected to be close to $2 \times k(1,0)$. Their interpretation is that the second smallest spatial frequency is $k(0,1)$. They deduce that the size of the Musca molecular cloud is $L_x \times L_y \sim 8.2 \times 6.2$ pc$^2$. They show that the next several low order modes fit the data well. Therefore, they claim that the Musca cloud is actually a rectangular-shaped very thin cloud that is oriented almost exactly edge on.

Since we have true filamentary clouds in our simulation and we observe striations as well, we can perform the same analysis as in \citet{tri18}. Note that we have constant turbulence driving in the entire simulated region instead of only the gravitational effect that was considered in the test models in \citet{tri18}. Nevertheless, it is of interest to find out how accurate the normal mode method is using our simulation results. We first obtain the $1024^2$ column density map at $0.5 \tff$ along the y-axis, as shown in Fig. \ref{striations_comp}a. We then select a stripe of size $100 \time 1024$ pixels above the filamentary cloud (white color box) that crosses over all the striation structures of the entire width of the simulated region. We also use the Lamb-Scargle periodogram technique (the Matlab function {\it plomb}) to perform spectral analysis of all the 100 lines ($1 \times 1024$ pixels) in the stripe. The peaks are identified using the Matlab function {\it findpeaks} and are added together from the 100 power spectra. The histogram is shown in Fig. \ref{striations_comp}b. The first peak corresponds to 4.6 pc, close to the projected length of the main cloud of 4.39 pc. However, the second and third strong peaks are not in exact multiple values of the spatial frequency of the first peak. The three vertical dashed lines in Fig. \ref{striations_comp}b are the exact multiples of the first peaks. Using the frequency of the second peak to work out the expected first peak, we have $k(1,0) = 0.84$ (instead of 0.683) corresponding to $L_x = 3.7$ pc, close to the projected length of the massive part of the main cloud. Also, the third peak matches very well with frequency $3\times k(1,0) = 2.52$. Treating the first peak as the (1,0) mode and the second peak as the (0,1) mode, we cannot find a peak that corresponds to (1,1) mode. Therefore, by using the first peak it is problematic to conclude that the three strong peaks are the first three normal modes. By ignoring the discrepancy of the first peak, these three peaks will be a good indication that the main cloud is a true filament because the corresponding $L_y$ of the cloud will be very small compared to $L_x$. The forth peak is unclear.  There are several peaks corresponding to $L_x \approx 0.7 \sim 1$ pc located next to each other around the expected locations of the forth and fifth peaks. This may be due to the presence of the secondary cloud with a projected width of about 1 pc on the map.

We have tried to move the stripe closer or further away from the main cloud and also below the cloud and the results are basically similar. There are variations in the histograms, but the first three strong peaks still exist. Closer to the cloud we note more noise in the form of several very small peaks in between. We then performed another analysis, but now using a shorter stripe. Instead of covering the entire map, we use $100 \times 500$ pixels to cover the visible striations (red color box in Fig. \ref{striations_comp}c). The reason for choosing a shorter stripe is because the stripe shown in the figure 1 in \citet{tri18} is not over the entire map or the entire length of the Musca cloud, but rather over the visible striations of width about 6 pc, which is shorter than the projected length of the Musca cloud. Our 500-pixel long stripe also covers the visible striations at that level. The histogram of this shorter strip is shown in Fig. \ref{striations_comp}c and it is quite different from that in \ref{striations_comp}b. The first and second peaks are marked with the spatial frequencies in the figure. The second peak is not a multiple of the first peak. Since the second peak has 10 counts, we cannot ignore it. The first peak corresponds to $L_x = 3.05$ pc, shorter than the projected length of the main cloud.  If we assume the second peak is coming from the dimension along the los, then $L_y = 2.18$ pc. The frequencies of the next several normal modes are shown by the vertical dashed-lines. The matching is not perfect given that the mode (2,0) is missing a peak and a peak at $k = 3.9$ does not have a corresponding normal mode and neither does figure 2B of \citet{tri18}. Based on the second analysis, one would misinterpret that there is a $3.05 \times 2.18$ pc$^2$ sheet-like cloud oriented almost edge on. When we vary the length of the stripe, the histogram varies as well and approaches the Fig. \ref{striations_comp}b when the stripe is long enough to cover mostly the entire width of the map. The second analysis result can also be due to the fact that there are two massive filamentary clouds in the T-shape configuration (see Fig. \ref{Nmaps}). The length of the secondary cloud length is about 2.5 pc.

From the above analysis, the width of the stripe used can affect the interpretation of the 3D geometry of molecular clouds using the normal mode analysis. Using our simulation data as an example, one can go from a long filamentary cloud to a 2D sheet-like structure, which in fact has two massive filamentary clouds in a T-shape configuration. The Musca molecular cloud could be a large sheet-like cloud as concluded in \citet{tri18} or it could be composed of more than one filamentary cloud that mimics a large sheet-like cloud from the normal mode analysis. Therefore, caution must be taken when interpreting the results from the normal mode analysis.

\begin{figure}
\includegraphics[scale=0.415]{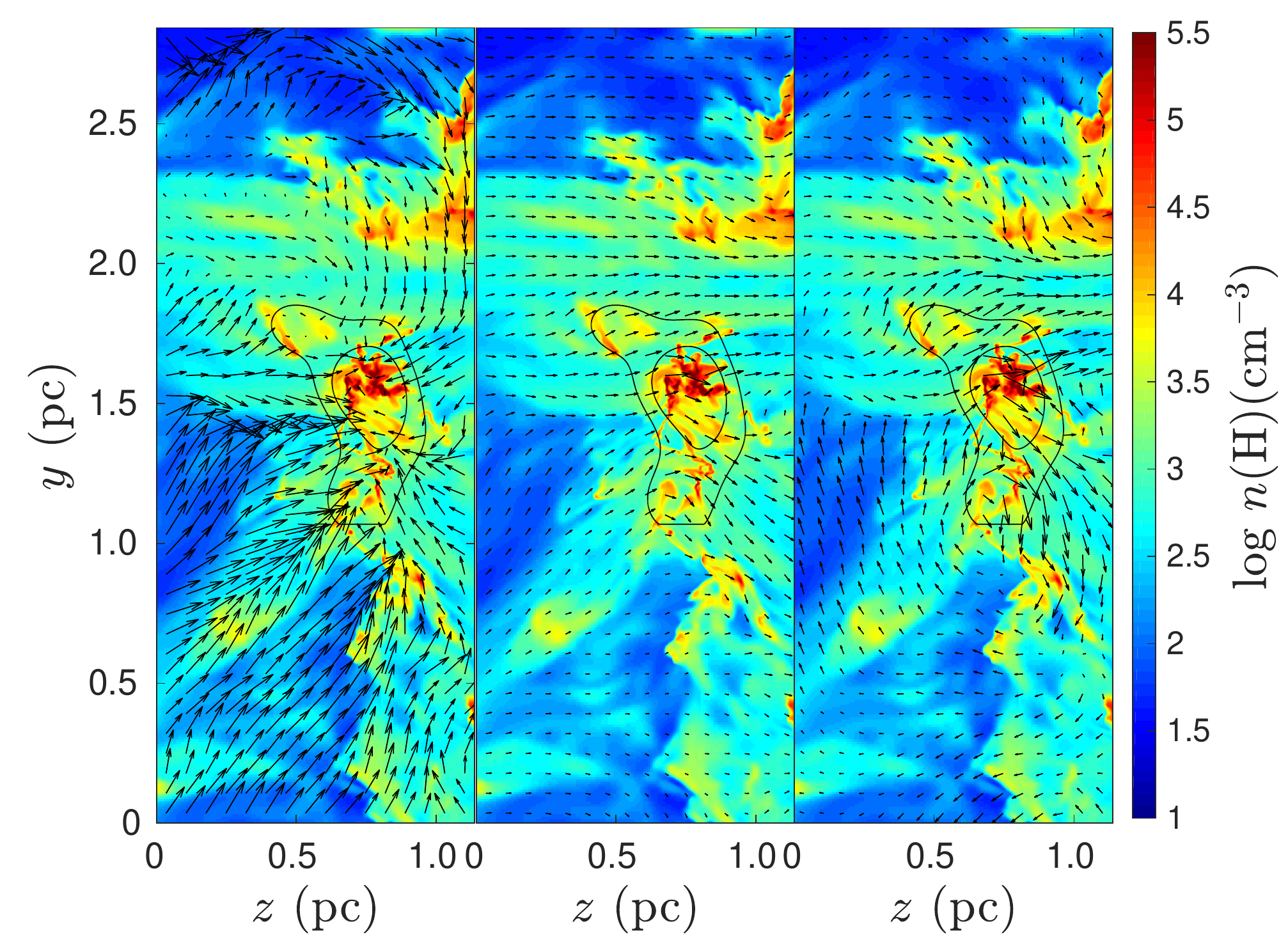}
\caption{Velocity (left), total magnetic field (middle), and turbulence magnetic fields (right) of a cross-section of the main cloud at time $0.5 \tff$. The turbulence magnetic field is the total field minus the mean magnetic field along the z-direction. The color map is the volume density of the cross-section. The two contours at log $n$(H) = 3.2 and 3.5 cm$^{-3}$, obtained from convolving the density map using a Gaussian beam of 0.1 pc radius, are plotted for visual identification of the location of the main cloud in the cross-section.
\label{bvNmaps}}
\end{figure}

\subsection{Magnetic field on the cross-section of the main cloud}
\label{sec:irdc_bxsect}
We have seen the large scale magnetic field structure around the main cloud in the above discussion. In Fig. \ref{bvNmaps}, we look more closely at the magnetic field and velocity field of a cross-section of the main cloud at $x = 2.22$ pc. Every cross-section along the main cloud is not exactly the same. We show the cross-section in Fig. \ref{bvNmaps} to describe several key common features of the density, magnetic field, and velocity field profiles. With our high resolution simulation, the main cloud density profile is far from a simple cylindrical profile but has a more complex substructure. The two contours at log $n$(H) = 3.2 and 3.5 cm$^{-3}$ are obtained by convolving the density with a 0.1 pc radius Gaussian beam. These two density contours in each panel indicate where the main cloud is located. In the panels, the magnetic field and velocity vectors are averaged over 32 finest cells at resolution 0.071 pc. The dense structures are the result of gravitational collapse and the shock compression from the converging flows predominantly along the magnetic field. The compression is not simply uni-directional due to the complex turbulent environment (left panel). As a result, the magnetic field is also not straightly piercing the cloud, but is twisted through the cloud (middle panel). Outside the dense cloud, the field is more or less closer to the mean field direction. Note that the mean field is along the z-axis, which is horizontal in the figure and we are looking along the positive x axis.
To see the morphology of the magnetic field due to turbulence, we subtract the total field by the mean field and plot the resulting field in the right panel in Fig. \ref{bvNmaps}. There is a toroidal field component not around but below the main cloud at this cross section as the result of the rotational and shearing flows of the turbulence. Separated toroidal field components appear at different locations in the simulated region after the subtraction. The toroidal field creates an inward force that helps squeeze the gas due to the magnetic tension. However, the spatial correlation of these separated short toroidal fields with the dense filamentary clouds generally is weak, with a spatial correlation coefficient less than 0.5. Nevertheless, we note that some of the toroidal field components are surrounding some fine filamentary substructures. The detailed study of velocity and magnetic fields of filamentary substructures will be discussed in paper III.

\section{Velocity structure}
\subsection{Large scale velocity structure}
\label{sec:irdc_lv}

\begin{figure}
\includegraphics[scale=0.93]{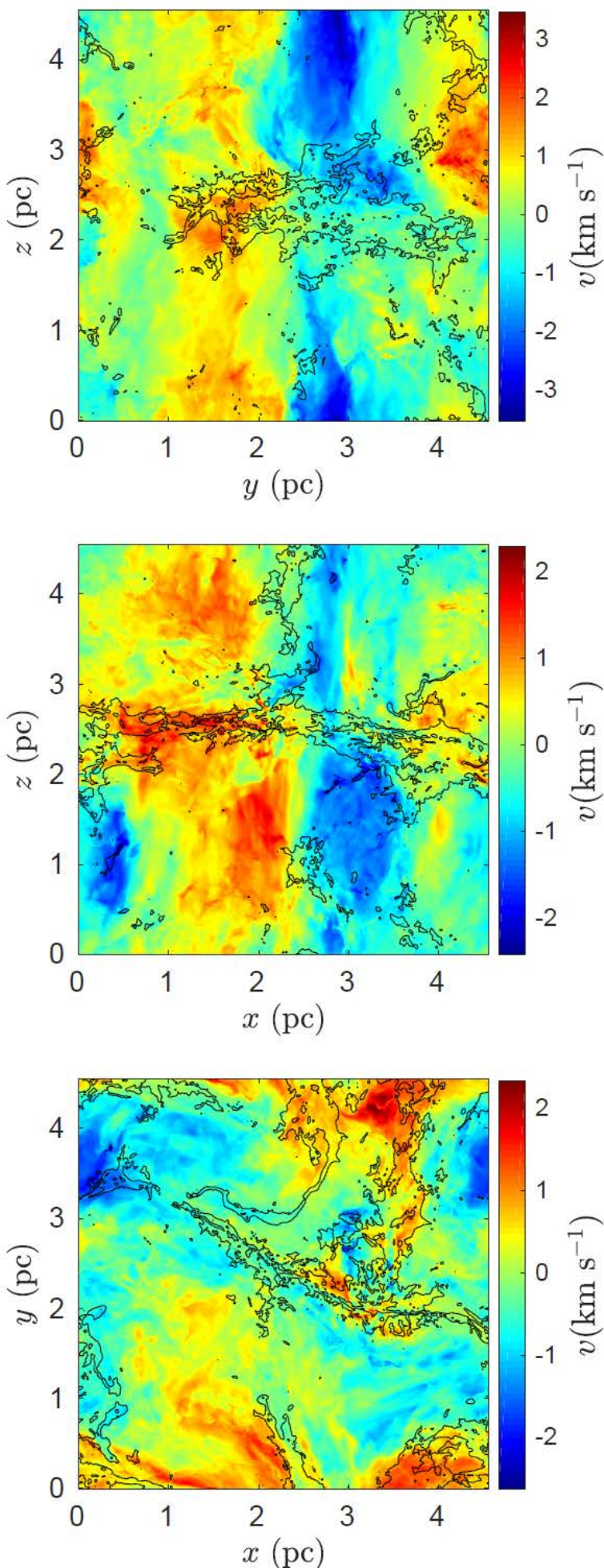}
\caption{Density-weighted los velocity maps along the x-, y-, and z-axis (top, middle, and bottom panel, respectively) of the entire simulated region. The contours are column density at $N({\rm H}) = 2$ and $6.3 \times 10^{22}$ cm$^{-2}$ at time $t = 0.5\tff$ as an outline of the filamentary clouds. The sign convention of the velocity is the same as that of spatial coordinates.
\label{vmaps}}
\end{figure}
Although Observations can obtain los velocity over a molecular cloud, the lack of a true 3D velocity field seriously limits our understanding of the spatial motion of gas at different parts of a filamentary cloud.  Observers sometimes rely on the los velocity gradient to interpret the 3D spatial movement of the gas.  This can result in an unreliable way of finding the spatial gas motion. In this section, we discuss the velocity structure of gas inside of and around the filamentary clouds obtained from the los velocity component in the simulation and explain the ambiguity that can arise by interpreting the 3D velocity structure using only the los velocity information.

In Fig. \ref{vmaps}, we show the density-weighted, integrated los velocity maps viewing along the three cardinal axes of the entire simulated region. The contours show the column density $N(H) = 2 \times 10^{22}$ and $6.3 \times 10^{22}$ cm$^{-2}$ at $t = 0.5 \tff$. We first consider the middle panel looking along the y-axis in the positive direction. The negative velocity points outward of the plane of the page. We note a large blue colored area near the middle and below the main cloud. The secondary cloud is located near the top tip of this blue color patch connected with the main cloud. At this time, the secondary cloud collides with the main cloud and deposits mass onto the main cloud. The direction of the merging secondary cloud gas points outward of the page. The projected area of the secondary cloud is quite small such that it is not easy to identify by just looking at the column density map. A velocity map will be helpful here.

\begin{figure}
\includegraphics[scale=0.48]{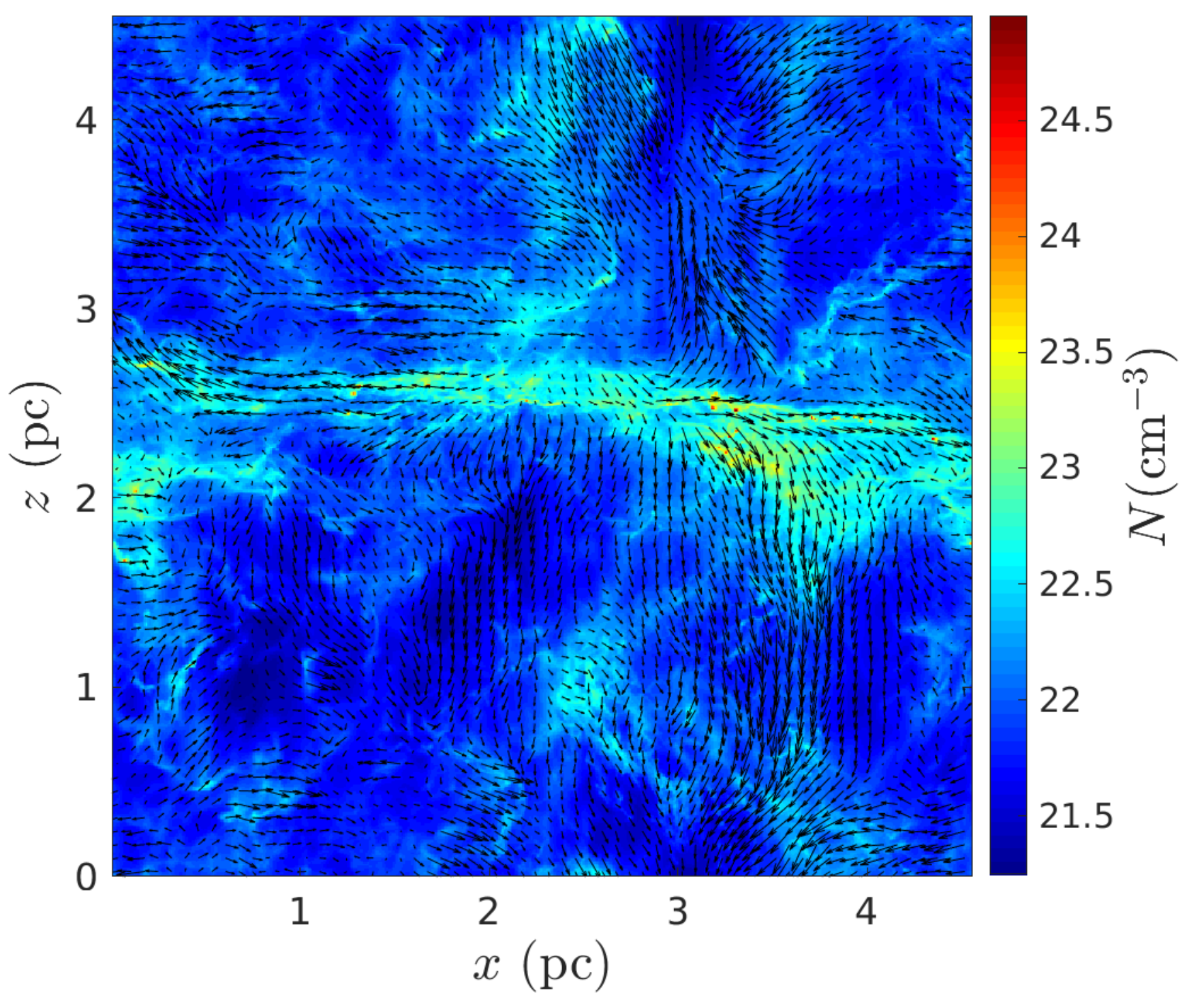}
\caption{Projected density-weighted velocity vectors on the column density map plane (x-z plane). The velocity vectors are at the resolution of 0.071 pc. The maximum velocity is 2.46 km s$^{-1}$.
\label{vmapxz}}
\end{figure}

We note a los velocity gradient along the main cloud in this projection in Fig. \ref{vmaps}. The gradient is small on the left side with los velocity close to 2 km s$^{-1}$ into the page and quickly changes to about -0.7 km s$^{-1}$, out of the page, near $x = 3$ pc. The los velocity slowly changes to close to 0.5 km s$^{-1}$ on the right end of the cloud. The change of los velocity from positive to negative and back to positive is simply because of the collision with the secondary cloud which moves the middle part of the main cloud to the direction out of the page. This is revealed by 3D information from the simulation (see the 3D volume rendering video). By relying on this 2D integrated map, one may conclude from the negative velocity gradient on the left side of the main cloud and the positive velocity gradient on the right side of the main cloud that the two sides of the cloud are merging together due to gravity. This is an incorrect conclusion. In fact, the density-weighted velocity on the plane of the map, shown in Fig. \ref{vmapxz}, shows that the cloud is slowly stretching. The stretching velocity gradient is $\sim 0.6$ km s$^{-1}$ pc$^{-1}$ at $0.5 \tff$. As previously mentioned in Section \ref{sec:cloudform}, the main cloud has never broken into two clouds due to continuous replenishment of gas moved to the cloud by large scale turbulence flow. From the above discussion, it is clear that the los velocity gradient provides information along the los only. One cannot use that to interpret the gas motion along the filament on the plane of the sky.

In the bottom panel of Fig. \ref{vmaps}, the los velocity of the main cloud along the z-axis also shows a velocity gradient from the two ends to the junction with the secondary cloud. The secondary cloud is moving in the direction out of the plane of the map (positive $v_z$) and collides with the main cloud. As a result of the collision, the gas of the main cloud at the junction is being pushed out of the map as well. In this direction, the velocity gradients and velocity dispersion of the gas of the two clouds are strikingly similar to the dark cloud SDC13 which will be discussed in Section \ref{sec:sdc13}. The top panel of Fig. \ref{vmaps} shows the los velocity of gas along the x-axis. In this projection, there is more velocity information of the secondary cloud, which is below the main cloud and connected at the ends at the left hand side of the map. Although the two clouds are spatially close in this projection, observations may still be able to distinguish two clouds from the quite different velocity gradients along the axes of the two clouds. The main cloud on top shows a larger gradient from tip to toe of $\sim 1.5$ km s$^{-1}$ pc$^{-1}$ but the secondary cloud on the bottom is about half of that. However, we may have other interpretations of the cloud gas movement if we do not know the true gas movement from the simulation. For example, one may combine the two clouds together as one and interpret that the gas on the right hand side is rotating about the y-axis. The argument is that the upper part, which is the main cloud, has a negative $v_x$ and the lower part has a positive $v_x$. Thus, it is non-trivial to understand the 3D spatial movement of the molecular cloud gas simply from the los velocity map.

\section{Physical properties of the main filamentary cloud}
\label{sec:cloudprop}

\begin{table}
\caption{Physical properties of the main cloud at $0.5 \tff$}
\label{tab:maincloud}
\begin{tabular}{lc}
\vspace{-0.3cm}\\
\hline
\hline
Length & 4.42 pc      \\
Mean width  & 0.25 pc \\
Mass   & $470.7 M_{\odot}$          \\
Mean density (n(H)) & $6.3\times10^4$ cm$^{-3}$  \\
Mean column density (N(H))\fnm[1] &  $3.8\times 10^{22}$ cm$^{-2}$\\
Mean magnetic field strength\fnm[1] & $51.9 ~\mu G$\\
Mean mass-to-flux ratio ($\mu_{\Phi}$)\fnm[1] & $2.78$\\
\hline
\end{tabular}
\begin{flushleft}
\fnt{1} {$^1$ Measured in projection along z-axis}\\
\end{flushleft}
\label{maincloud}
\end{table}

Here we summarize the physical properties of the main cloud. Since the secondary cloud disintegrates at later time, we only provide the physical properties of the long lasting main cloud at $0.5 \tff$. From the evolutionary picture depicted in Fig. \ref{Nevol}, the appearance and physical size of the main cloud are changing in time. As expected, the total mass, mean density, and maximum density are all functions of time. In addition, the 3D geometry of the filamentary cloud is clearly not a simple cylinder, even though a cylindrical model is usually used to describe a filamentary cloud. In fact, it is more accurate to describe a long filamentary cloud as a collection of fiber-like substructures \citep[e.g.][Li, McKee, \& Klein 2019, in preparation]{hac13}. For discussion purpose in this section, we still use the highly simplified cylindrical geometry to describe the mean physical properties of the main cloud listed in Table \ref{maincloud}. To define the extent of the main cloud, we use the column density maps from the three cardinal axis projections. Generally, an IRDC is defined to have mean column density of $N({\rm H_2}) \approx 10^{22}$ cm$^{-2}$. Therefore, we define the envelope of the cloud to be when the column density $N({\rm H}) = 2 \times 10^{22}$ cm$^{-2}$. By this definition, we can determine the area of the cloud on each projection. The length of the cloud is measured by dividing the main cloud into many short sections of roughly 0.1 pc long along the long axis. We find the maximum density of these sections and define them as the centers along the filamentary cloud. The sum of the distances between each pair of section centers will give us the length of the cloud. By using this method, the 3D length of the cloud is found to be about 4.42 pc at $0.5 \tff$. We use the same procedure to measure the projected length of the main cloud in each cardinal projection and compute the mean width of the cloud. The mean width varies between 0.19 to 0.36 pc from the three projections, similar to the width of filamentary clouds such as SDC13 and L1495. The average width of the cloud is $\sim 0.25$ pc at this time. The mean volume density and projected column density of the main cloud at this time are determined by the total mass within the envelop of the main cloud and listed in Table \ref{maincloud}.

We measure the mean mass accretion rate using the total mass within the envelop of the main cloud between 0.3 and $0.5 \tff$. Using the above definition of the extent of the main cloud, the mass of the main cloud is $415.7 M_{\odot}$ and $470.7 M_{\odot}$ at times $0.3 \tff$ and $0.5 \tff$, respectively. If we define the mass accretion rate to be material accreted into the contour of $N({\rm H}) = 2 \times 10^{22}$ cm$^{-2}$, the mean accretion rate of the main cloud over this time period is $1.96 \times 10^{-4} M_{\odot} {\rm yr}^{-1}$. If we use the length of the main cloud at $0.5 \tff$ of 4.42 pc, it implies that the accretion rate of the main cloud is $\sim 44.3 M_{\odot}~{\rm Myr}^{-1} {\rm pc}^{-1}$. For comparison, the B211 dark cloud has an estimated accretion rate of $27 - 50 M_{\odot}~{\rm Myr}^{-1} {\rm pc}^{-1}$ \citep{pal13}. If we measure the mean accretion rate up to $0.64 \tff$, which will include the effect of the large amount of gas merged from the secondary cloud to the main cloud due to the collision, the mean accretion rate of the main cloud is $72.8 M_{\odot}~{\rm Myr}^{-1} {\rm pc}^{-1}$. 

The entire main cloud has little movement. The center of the cloud displaces only about 0.2 pc and the cloud rotates slowly by $\sim 26^{\circ}$ during the entire simulation. This is probably one of the reasons that the long filamentary cloud can last for the entire duration. The mean mass-to-flux ratio, $\mu_{\Phi} \sim 2.78$, of the main cloud in Table \ref{maincloud} is measured along the mean field direction (z-axis) over the area defined above in determining the column density of the main cloud. Note that initially the $\mu_{\Phi}$ of the entire simulated domain is 1.62, as shown in Table \ref{tab:region}. In \citet{li15}, the measured median $\mu_{\Phi}$ of dense clumps at this time is 2.63, similar to the mean value of the main cloud.

\section{Comparison with the dark cloud SDC13}
\label{sec:sdc13}
\begin{table}
\caption{Physical Properties of the filamentary clouds in SDC13 and from the simulation}
\label{tab:sdc13}
\begin{tabular}{lcc}
\vspace{-0.3cm}\\
\hline
\hline
& SDC13 & simulation \\
\hline
Total mass\fnm[1] ($M_{\odot}$) & 1276 & 741 \\
Mass\fnm[1] of the main cloud ($M_{\odot}$) & 616\fnm[2] & 470.7 \\
Length of the main cloud (pc) & $\sim 4.23$\fnm[2] & $\sim4.39$\fnm[3] \\
Mass per unit length of \\
~~~the main cloud\fnm[4] ($\times 2c_s^2/G$) & 7.4 & 6.5 \\
Mass of the most massive \\
~~~clump\fnm[5] ($M_{\odot}$) & $\sim 80$ & $\sim 78$ \\
Width of all clouds (pc) & $0.16\sim0.37$ & $0.25\sim0.35$ \\
Mean column density $N$(H) of \\
~~~all clouds ($\times10^{22}$ cm$^{-2}$) & 5.70 & 3.61\fnm[6] \\
\hline
\end{tabular}
\begin{flushleft}
\fnt{1} {$^1$ The mass measured from $N({\rm H})\ge2\times10^{22}$ cm$^{-2}$.}\\
\fnt{2} {$^2$ The main cloud in SDC13 is assumed to be Fil-NW + Fil-SE.}\\
\fnt{3} {$^3$ Projected length on the x-y plane.}\\
\fnt{4} {$^4$ In unit of thermal critical mass per unit length $2c_s^2/G$. The sound speed in SDC13 is based on temperature of 12K and that in the simulation is based on 10K.}\\
\fnt{5} {$^5$ Both clumps are at the junction of the filaments in SDC13 and in the simulation.}\\
\fnt{6} {$^6$ Only the main and secondary clouds in the simulated region.}\\
\end{flushleft}
\label{sdc13}
\end{table}

The column density map at the right panel of the Fig. \ref{Nmaps} shows a T-shape configuration which is remarkably similar to the IRDC SDC13 \citep{per14,wil18}. From the right panel of Fig. \ref{Nmaps}, there is a lower density reversed-C-shaped gas stream near the left side and above of the main cloud. This gas stream has column density barely larger than $2 \times 10^{22} {\rm cm}^{-2}$. A part of this reversed-C-shaped gas stream moves to the direction of the main cloud and becomes gravitationally bound to the main cloud at a later time. We do not include this gas stream to the main cloud at $0.5 \tff$, when we measure the total mass of the main cloud. Although the detailed gas distribution around SDC13 and our simulated IRDCs are not exactly the same, the T-shaped configuration, where most of the gas is concentrated in SDC13 and in the two filamentary clouds in our simulation, are similar. The spatial extent of SDC13 is also similar to the T-shape IRDCs in the simulation. The combined length of Fil-SE and Fil-NW in SDC13 is about 4.23 pc, similar to the projected length of 4.39 pc of the main cloud in the simulation. The length of the Fil-NE is 2.89 pc and the length of the secondary cloud in the simulation is about 2.5 pc. The total mass of SDC13 above $N(H) > 2 \times 10^{22} {\rm cm}^{-2}$, including all four filaments is $1276 M_{\odot}$. The total mass of the main and secondary clouds in the simulation is $\sim 58$ per cent of that. In Table \ref{sdc13}, we list the comparison of the T-shaped IRDCs from our simulation with SDC13. In \citet{per14}, they discuss the gas flow along the filaments based on the los velocity and conclude that the junction of the filaments is where all the gas is converging. There are some very massive clumps near the junction and the most massive is $80 M_{\odot}$. These massive clumps are expected to be formed as the result of gas collapsing onto the junction. As reported in \citet{li15}, the most massive clump in our simulation is about $78 M_{\odot}$ and is also located at the junction of the main and secondary clouds where they collide. Since our simulated filamentary IRDC is so similar to SDC13, we determine if we have a similar gas flow pattern as SDC13 based on los velocity.

\begin{figure}
\includegraphics[scale=0.44]{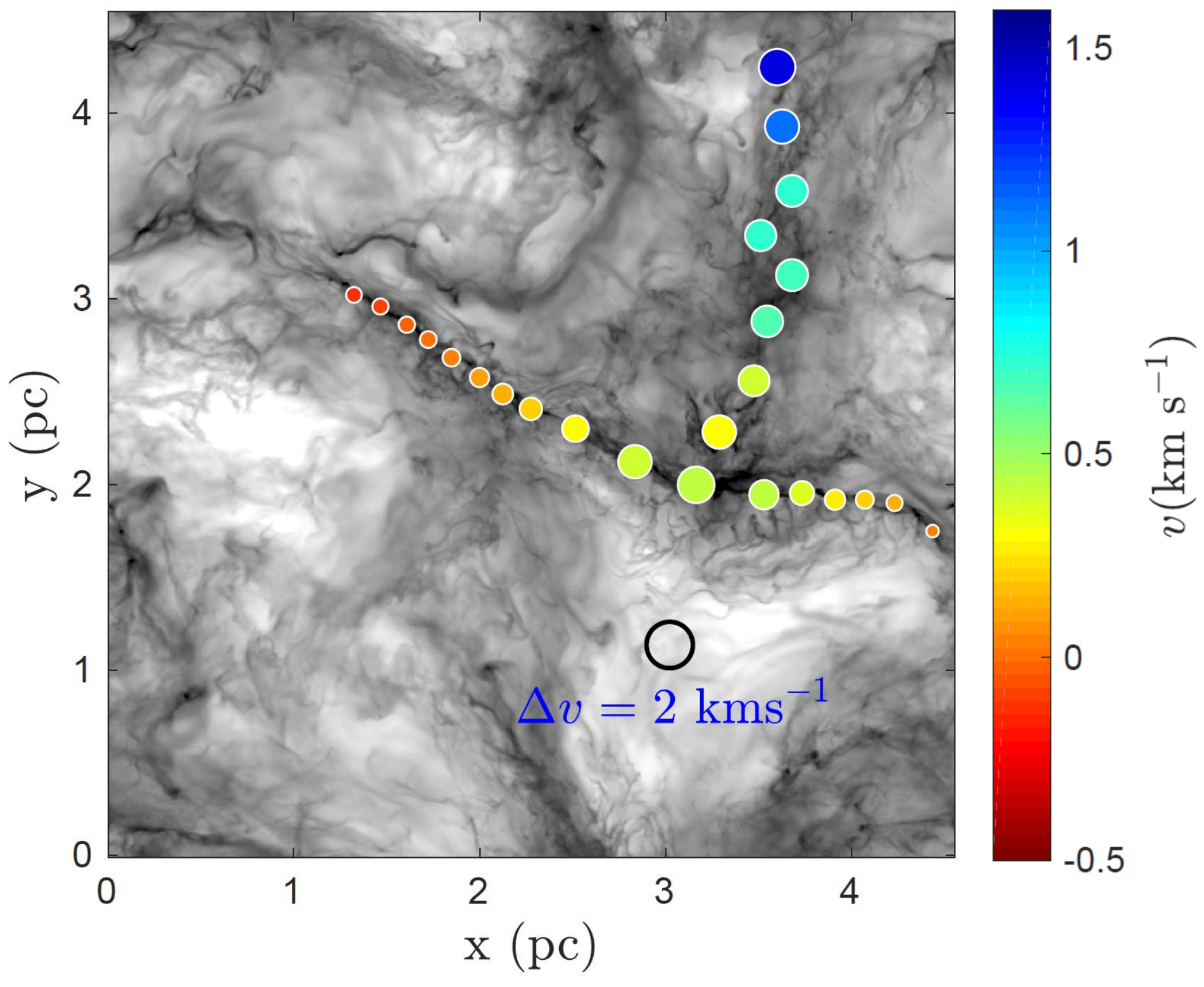}
\caption{Projected column density map along the z-axis in grey scale with los velocities (in color) and velocity dispersion (represented by the size of the circles). A circle of velocity dispersion 2 km s$^{-1}$ is shown on the map as a reference. See Section \ref{sec:sdc13} for discussion.
\label{vdisp}}
\end{figure}
\begin{figure*}
\includegraphics[scale=0.81]{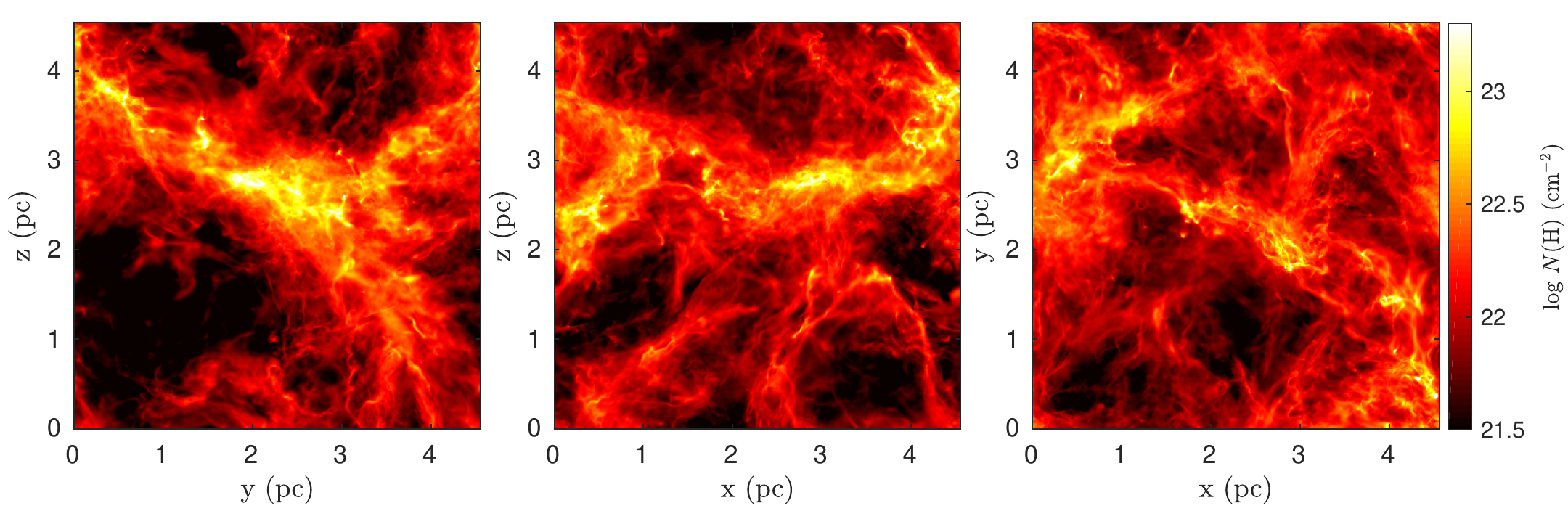}
\caption{Column density maps of the initial weak magnetic field model viewing along the three cardinal axes at time $t = 0.5 \tff$. Dark clouds are clumpy compared to the initial strong magnetic field model in Fig. \ref{Nmaps}.
\label{wNmaps}}
\end{figure*}

We create a map similar to figure 4 in \citet{per14} for the comparison of the los velocity gradients along the filamentary clouds. The observed beam size of the 31 points using N$_2$H$^+$(1-0) for measuring the los velocity in \citet{per14} is $27''$, corresponding to 0.47 pc. From our simulation, we first create a density-weighted los velocity map along the z-axis (see the bottom figure in Fig. \ref{vmaps}) and compute the velocity dispersion for every pixel. We then use a Gaussian beam of 0.47 pc to convolve the velocity and velocity dispersion maps. The los velocity and velocity dispersion are measured at 25 points along the main and secondary clouds and are shown in Fig. \ref{vdisp}. From this figure, the largest velocity dispersion is less than 2 km s$^{-1}$, similar to SDC13. The colors of the markers along the main cloud change in a similar way as SDC13. For the main cloud, the two ends have greater negative velocity than near the cloud junction. For the secondary cloud, the end near the junction is more negative in velocity than the other end. From the velocity gradient, we can conclude as in SDC13, that all three legs of the T-configuration clouds are coalescing at the junction. As shown in Fig. \ref{vmapxz}, the main cloud is not systematically collapsing onto the junction, but is slowly stretching at a mean velocity gradient of $\sim 0.6$ km s$^{-1}$ pc$^{-1}$ at $0.5 \tff$. This shows that interpreting 3D spatial gas flow based only on a los velocity gradient may lead to a very different conclusion. In the simulation, the gradient of los velocity in Fig. \ref{vdisp} is the result of a collision between the main and the secondary clouds. As discussed in Section \ref{sec:irdc_lv}, the main cloud moves into the map and the secondary cloud moves out of the map. The central part of the main cloud collides with the secondary cloud and is slowed down. Similarly, the end of the secondary cloud near the interaction is also slowed down by the main cloud due to the collision. This results in the los velocity gradients seen in Fig. \ref{vdisp}. This interpretation may be equally applied to the observed los velocity gradient of SDC13. Lack of knowledge of the gas velocity on the plane of sky, prevents us from knowing the true 3D gas flow directions along the filamentary clouds in SDC13.

In \citet{wil18}, the los velocity maps of SDC13 using NH$_3$(1,1) and NH$_3$(2,2) show large los radial velocity gradients up to 3 km s$^{-1}$ pc$^{-1}$ in Fil-NE and Fil-SE. Their analysis concludes that the large radial velocity gradient is due to compression from nearby star formation activity. In our IRDCs simulation, we do not yet have such star formation activity simulated nearby and we do not see large radial velocity gradients near the ends of the main cloud. However, we do see large los radial velocity gradients of $2 \sim 3$ km s$^{-1}$ pc$^{-1}$ near the junction, which is due to the collision of gas flows of the two clouds, as well as near the other end of the secondary cloud, which is the result of large shearing and rotation of turbulent gas around the less collimated secondary cloud. Another striking similarity between the SDC13 and the IRDCs in our simulation is the morphology of the junction. A thin but long filament ``N'' is identified between Fil-NE and Fil-NW in the SDC13 in \citet{wil18}. The junction location of all the filaments in SDC13 shows a radial-like appearance. We see a similar appearance in the bottom panel of Fig. \ref{Nmaps}. In the process of collision, some short fiber-like gas streams from the secondary cloud also are pointing roughly radially to the junction and merging with the main cloud. Together with the branching of the main cloud on the right hand side of the junction, they form a radial-like morphology.

\section{What makes long filamentary IRDCs?}
\label{sec:irdc_form}
\begin{figure*}
\includegraphics[scale=1]{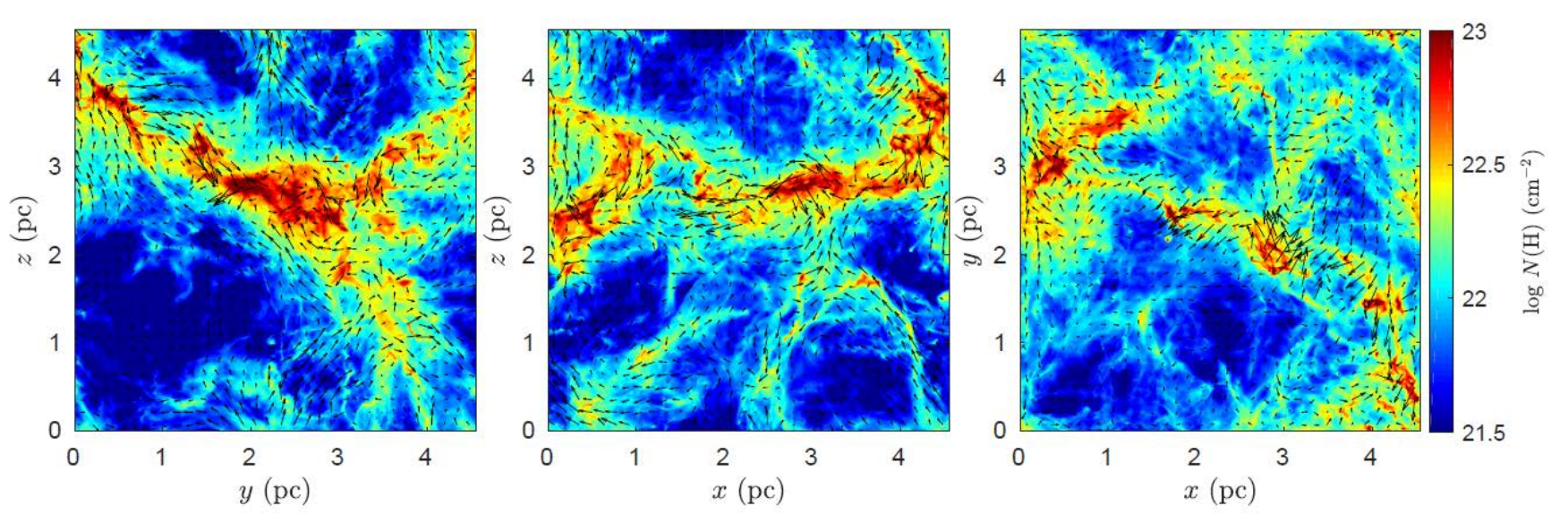}
\caption{The los density-weighted magnetic field, on top of the column density map, along the x-, y-, and z-axis (left, middle, and right panel, respectively) of the entire simulated region from a initially weak magnetic field model. The magnetic field orientation is highly random in all three projections. The first two panels are very different from Fig. \ref{bfield}.
\label{weakbfield}}
\end{figure*}
\begin{figure*}
\includegraphics[scale=0.8]{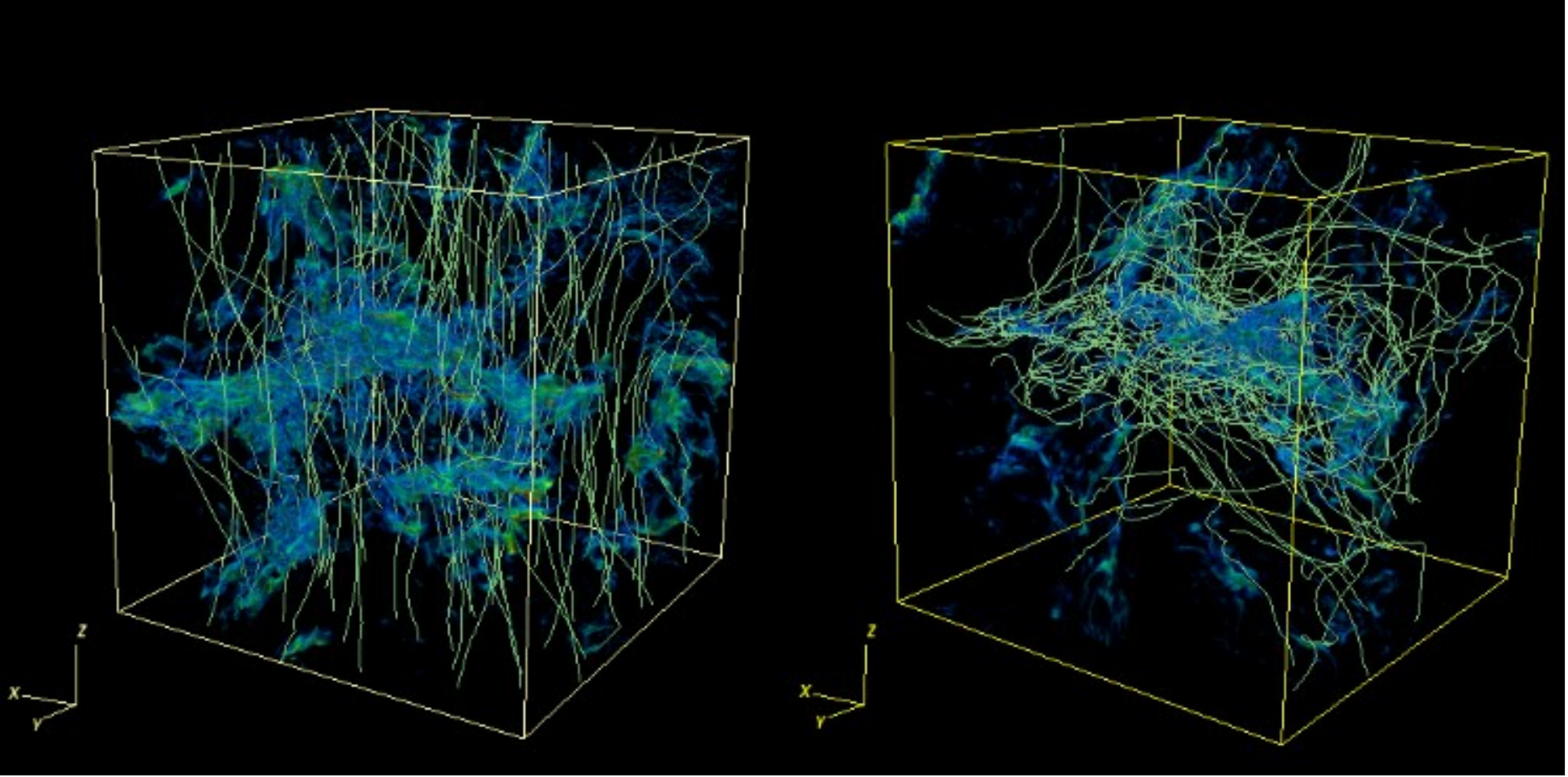}
\caption{Comparison of magnetic field streamlines in the strong field model with initial $\ma=1$ (left) and in the weak field model with initial $\ma=10$ (right) at $0.5 \tff$.
\label{bstream_comp}}
\end{figure*}

Why do filamentary dark molecular clouds have a long filamentary shape and how does the filamentary cloud remain long and slender? We address the questions by comparing two simulations here. Long and slender filamentary clouds are not rare. That suggests that the conditions for the formation of long filamentary clouds may be quite common in GMCs. We have performed two simulations in our study of IRDCs. The comparison of the magnetized dense clumps formed in the two simulations were first discussed in \cite{li15}. The two simulations have exactly the same initial conditions, including the turbulence driving pattern. They only differ in the initial magnetic field strength. The strong field model has an initial \alfven Mach number of one and the weak field model has an initial \alfven Mach number of ten, hence a ten times difference in the initial magnetic field strength. We have discussed the detailed results from the strong field model in this paper which showed the formation of a long filamentary cloud of length 4.42 pc, similar to a typical long filamentary dark cloud. To illustrate comparison with Fig. \ref{Nmaps} we show the column density images of the weak field models at time $0.5 \tff$ in Fig. \ref{wNmaps}.  The difference is clear that the massive dark clouds in the weak field model are more clumpy.

Since the only difference between these two models is the initial magnetic field strength, clearly a necessary condition for forming a long and slender filamentary cloud is the presence of a moderately strong magnetic field. In Fig. \ref{weakbfield}, the density-weighted magnetic field in the initially weak field model is highly random in all three projections. In Fig. \ref{bstream_comp}, we plot the streamlines of the magnetic field in the volume rendering of density of these two simulations at $0.5 \tff$. The magnetic field in the strong field model is strong enough to resist the large perturbations from turbulence and remains closely aligned along the z-axis. The less perturbed magnetic field acts as a reinforcement to hold the long filamentary cloud gas together without buckling or breaking. In the weak field model, the magnetic field strength is not strong enough to resist the stretching due to turbulence flow. Therefore, gas is unable to remain in a long filamentary form, but tends to form smaller dense clumpy structures. From observations, the \alfven Mach number of molecular clouds is close to unity. Finding long filamentary dark clouds in GMCs should not be difficult.

From our simulations, we conclude that in order to form a massive long filamentary cloud, strong magnetic fields play a crucial role in guiding the gas flow and reinforcing the long filamentary structure. Large scale turbulent flow plays the role of sweeping the gas together and gravity plays the role of keeping the gas together.

\section{Conclusions}
\label{sec:conclusion}
Massive filamentary dark clouds have received a lot of attention lately, in part due to the Herschel space telescope that reveals fine fiber/web-like substructures and spectacular chains of cores forming along filamentary clouds. With additional information on magnetic fields and kinematic information, observers have started to assemble a picture of how filamentary clouds form in the first place and how turbulence and magnetic fields affect the formation process. However, due to limited information from the los velocity field and the 2D projected magnetic fields on the plane-of-sky, it is a challenge to fully decipher the true 3D spatial configuration of dark clouds, their large and small scale motion, and the magnetic field structure. Speculations on the true structure of dark clouds and their formation are unavoidable. Numerical simulations are crucial to understand the structure and formation of filamentary dark clouds. Due to the heavy computational cost of driven turbulence MHD simulations, current numerical studies on filamentary clouds either do not include magnetic fields or the MHD simulations do not cover a large enough dynamic range in spatial resolution. A scarcity of information is reported on magnetic fields from those simulations even when the magnetic field is included. In this paper, we report our AMR high resolution results on the formation of filamentary IRDCs covering a (4.55 pc)$^3$ region that allows the formation of clouds more than 4 pc in length and a resolution that achieves $\sim  2.2\times 10^{-3}$ pc, which is high enough to reveal fiber/web-like substructures inside filamentary clouds.

We have performed two simulations investigating the formation of filamentary IRDCs. They have exactly the same initial conditions except for a factor of ten in the initial mean magnetic field strength. Massive long filamentary IRDCs are formed in the initially strong field model and the IRDCs in the initially weak field model are clumpy. Our analysis in this paper mainly focuses on the strong magnetic field model. A 4.42 pc long,
$\sim 450 M_{\odot}$ slender filamentary cloud is formed and remains intact up to 2/3 of a free-fall time ($\sim 900,000$ yrs) in the simulation. The length of the main cloud is close to the size of the simulated region at the end of the simulation.  However, for a simulation with periodic boundaries, a filamentary cloud that is formed in the simulation could be larger than 4.55 pc depending on the orientation and the shape.
The global physical properties of the main cloud are listed in Table \ref{tab:region}. A nearby less massive filamentary cloud formed at about the same time collides with the main cloud around $0.43\tff$, dumping a large amount of gas onto the main cloud. During the collision, these two massive filamentary clouds form a T-shape configuration that resembles the dark cloud SDC13 \citep{per14}. We compare the large scale structures and properties of the dark clouds in the simulation with SDC13 in Section \ref{sec:sdc13}. After the collision, the secondary cloud breaks into three pieces at later time. The main cloud remains intact.

With the time evolution from our simulations, full 3D information on velocity, density, and magnetic field over 2/3 of a free-fall time is obtained and we are able to observe the filamentary clouds in different directions that observations are not able to do. With this information, we are able to demonstrate how relying on the los velocity or the plane-of-sky magnetic field information can result in misleading interpretations. In this paper, we report the results of the simulation and comparison with observations mainly on the large scale properties of the filamentary IRDCs. Our findings are summarized below:

\begin{itemize}

\item[1.] {\it Time evolution of the turbulence system during gravitational collapse.} When gravity is turned on in a self-consistent driven MHD turbulence system at the equilibrium state, the density power spectrum inertial range power law index evolves quickly from negative to positive. It is the result of gravitational collapse of gas at small scales forming dense clumps and cores. The high density end of both the volume and column density PDFs evolves from a log-normal distribution to a power law. The power law indexes increases to about -1.49 in the volume density PDF and about -2 in the column density PDF at the end of the simulation. This range is within the range of power law indexes of observed dark clouds \citep[e.g.][]{lin16,lin17}.
From our simulations, the differences from observations in power law indexes of dark cloud column density may indicate different epochs of the gravitational collapse phase. The velocity spectrum inertial range stays constant with power law index $\sim -1.54$ throughout the entire gravitational collapsing phase. However, the power at the high-wavenumber end is slowly increasing as the result of the conversion of gravitational energy into kinetic energy at small scales. The density structure of the filamentary clouds in the simulation is complex and non-monolithic. Using the algorithm {\it getfilament} \citep{men13}, complex fiber/web-like substructures are identified within the filamentary dark clouds. If the observational resolution is low, the complex substructures can be washed out and appear as a single filament using these algorithms. The detailed discussion of the small scale density structure of filamentary dark clouds will be presented in paper III.

\item[2.] {\it The role of the large scale strong magnetic field in the formation of long filamentary clouds.} Self gravitating filamentary dark clouds in the driven turbulence simulation are constantly subjected to large scale supersonic turbulence forcing. The main cloud remains intact for the entire time due to both the small spatial movement of the entire cloud with respect to the mean magnetic field and that the strong field acts as the reinforcement to hold the long filamentary cloud together much longer without breaking or buckling. The strong field also acts as a guiding channel for the gas to flow along the field directly onto the main cloud. In the initially weak field model, the magnetic field is too weak to withstand the shearing force from the turbulence flows and becomes tangled. As a result, massive dense gas is coalesced into clumpy structures instead of a long and slender filamentary cloud.

\item[3.] {\it Magnetic properties of the filamentary cloud.} For the long lasting main cloud, the large scale magnetic field pierces the cloud roughly normal to the cloud's long axis, similar to the field orientation around massive long dark clouds seen in polarization observations \citep[e.g.][]{cha11,pal13}. Depending on the gas motion near the cloud, the magnetic field can be substantially bent at some of the locations along the cloud. This can lead to an interpretation suggesting a helical structure of the magnetic field from Zeeman observations \citep{hei87} at some viewing directions.  From our study of the clumps formed in the cloud (\citet{li15}) we find that the mean mass-to-flux ratio of the main cloud is $\sim 2.78$, consistent with the median mass-to-flux ratio 2.63 of the dense clumps, which are located along the filament. The magnetic field structures at small scales inside the filamentary cloud are much more complex and will be discussed in paper III.

\item[4.] {\it Projection effect on magnetic field map.} Using our simulations, we demonstrate that a projected magnetic field on the plane-of-sky can be significantly different from the actual 3D magnetic field structure as obtained in the simulations. Because of the moderately strong mean magnetic field used in the simulation, the magnetic field is mostly aligned along the mean field direction, but has larger perturbations inside the dense clouds. Therefore, the projected large-scale field from most viewing angles will be roughly normal to the filamentary dark cloud's long axis. However, if the viewing direction is close to the mean field direction, the projected plane-of-sky field will be mostly from the random turbulence excited magnetic field that is perpendicular to the mean field. The field strength measured using the DCF method on the plane-of-sky in this case will be very small and lead to the conclusion that the magnetic field is unimportant throughout the entire region.

\item[5.] {\it Striations around filamentary clouds.} We can visually identify striation structures around the filamentary clouds and they are well matched with the large scale magnetic field. This gives strong support for the existence of a large scale relatively strong magnetic field around filamentary clouds implied from observations \citep{pal13}. We have examined the method of normal modes of striations around filamentary clouds \citep{tri18} to determine the dimensional structure of dark clouds in the simulation and we find the method can lead to different conclusions dependent on the window size covering the striations.

\item[6.] {\it Velocity structures.} 
The main cloud is formed and remains around the converging location of the large scale turbulence flows. The secondary cloud is spatially moving towards and collides with the main cloud. The entire main cloud has very slow spatial movement. Using the density-weighted los projected velocity maps, the longitudinal velocity gradients of the two clouds are about $0.5 \sim 1$ km s$^{-1}$ pc$^{-1}$, similar to the observed ranges of velocity gradient \citep[e.g.][]{per14,wil18}. The los velocity gradient in the main cloud reverses direction due to a collision at the junction with the secondary cloud. The magnitudes of the los velocity gradients and velocity dispersions along the filamentary clouds are similar to those of IRDC SDC13 \citet{per14}. SDC13 may be interpreted as a collision of two long massive filamentary clouds. The main cloud in the simulation stretches longitudinally due to large scale solenoidal turbulence driving as gas is continuously replenished from the large scale turbulence flows so that the main cloud does not break into two parts.

\item[7.] {\it Key factors on forming long and slender filamentary clouds.}
Our simulations show that the three key factors in making a long filamentary clouds are: (1) large scale converging turbulence flows that bring the shock-compressed material together, (2) gravity that holds dense gas together without further dispersion by supersonic turbulence and forms denser structures in the cloud, and (3) moderately strong magnetic fields that restrain the flow of dense gas and reinforce the appearance of the long cloud structure.

\end{itemize}

The scenario we derive from our simulations of the formation of long filamentary clouds is based upon one set of initial conditions and may not be able to explain all the filamentary clouds observed. Further simulations with different initial conditions are necessary in order to understand the different shapes and dynamics of dark filamentary clouds.

\noindent
\section*{Acknowledgments}
We thank an anonymous referee for helpful comments and suggestions for the paper. We also thank Chris McKee and Tie Liu for their helpful comments and suggestions on this paper. PSL would like to thank Alexander Men'shchikov on how to use the code \textsc{getsources/getfilament}. Support for this research was provided by NASA through a NASA ATP grant NNX17AK39G (RIK \& PSL) and the US Department of Energy at the Lawrence Livermore National Laboratory under contract DE-AC52-07NA 27344 (RIK).  This work used computing resources from an award from the Extreme Science and Engineering Discovery Environment (XSEDE), which is supported by National Science Foundation grant number ACI-1548562, through the grant TG-MCA00N020, computing resources provided by an award from the NASA High-End Computing (HEC) Program through the NASA Advanced Supercomputing (NAS) Division at Ames Research Center, and an award of computing resources from the National Energy Research Scientific Computing Center (NERSC), a U.S. Department of Energy Office of Science User Facility operated under Contract No. DE-AC02-05CH11231.

\label{lastpage}

\end{document}